\pdfoutput=1

\documentclass[11pt,twoside,a4paper,cmspaper,final,collab]{cms-tdr}
\def\svnVersion{455442P}\def\cmsCernNoTag{CERN-EP-2017-283}\def\cmsCernDate{\today}\def\cmsMessage{Published in the Journal of High Energy Physics as \href{http://dx.doi.org/10.1007/JHEP03(2018)160}{\doi{10.1007/JHEP03(2018)160.}}}

\begin{document}\cmsNoteHeader{SUS-17-004}

\hyphenation{had-ron-i-za-tion}
\hyphenation{cal-or-i-me-ter}
\hyphenation{de-vices}
\RCS$Revision: 452019 $
\RCS$HeadURL: svn+ssh://svn.cern.ch/reps/tdr2/papers/SUS-17-004/trunk/SUS-17-004.tex $
\RCS$Id: SUS-17-004.tex 452019 2018-03-21 14:59:06Z cheidegg $

\newcommand{\fullLumi}{35.9\fbinv}
\newcommand{\secondchi}{\PSGczDt}
\newcommand{\firstchi}{\PSGczDo}
\newcommand{\firstcharg}{\PSGcpmDo}
\newcommand{\gravitino}{\PXXSG}
\newcommand{\chargino}{\PSGcpm}
\newcommand{\neutralino}{\PSGcz}
\newcommand{\chargneut}{\ensuremath{\firstcharg\secondchi}\xspace}
\newcommand{\neutneut}{\ensuremath{\firstchi\firstchi}\xspace}
\newcommand{\firstchargpair}{\ensuremath{\firstcharg\widetilde{\chi}^{\mp}_{1}}\xspace}
\newcommand{\msecondchi}{\ensuremath{m_{\PSGczDt}}\xspace}
\newcommand{\mfirstchi}{\ensuremath{m_{\PSGczDo}}\xspace}
\newcommand{\mfirstcharg}{\ensuremath{m_{\PSGcpmDo}}\xspace}
\newcommand{\bfchih}{\ensuremath{\mathcal{B}(\firstchi\to\PH\gravitino)}\xspace}
\newcommand{\wz}{\ensuremath{\PW\PZ}\xspace}
\newcommand{\wh}{\ensuremath{\PW\PH}\xspace}
\newcommand{\zz}{\ensuremath{\PZ\PZ}\xspace}
\newcommand{\zh}{\ensuremath{\PZ\PH}\xspace}
\newcommand{\hh}{\ensuremath{\PH\PH}\xspace}
\newcommand{\ttw}{\ensuremath{\ttbar\PW}\xspace}
\newcommand{\ttz}{\ensuremath{\ttbar\PZ}\xspace}
\newcommand{\WJ}{\ensuremath{\PW\mathrm{+jets}}\xspace}
\newcommand{\MuMu}{\ensuremath{\Pgmp\Pgmm}\xspace}
\newcommand{\ElEl}{\ensuremath{\Pep\Pem}\xspace}
\newcommand{\EM}{\ensuremath{\Pe^\pm\Pgm^\mp}\xspace}
\newcommand{\hgg}{\ensuremath{\PH(\cPgg\cPgg)}\xspace}
\newcommand{\mll}{\ensuremath{m_{\ell\ell}}\xspace}
\newcommand{\MT}{\ensuremath{M_\text{T}}\xspace}

\cmsNoteHeader{SUS-17-004}
\title{Combined search for electroweak production of charginos and neutralinos in proton-proton collisions at $\sqrt{s} = 13\TeV$}

\date{\today}

\abstract{
A statistical combination of several searches for the electroweak production of charginos and neutralinos is presented. All searches use proton-proton collision data at $\sqrt{s}=13\TeV$, recorded with the CMS detector at the LHC in 2016 and corresponding to an integrated luminosity of 35.9\fbinv. In addition to the combination of previous searches, a targeted analysis requiring three or more charged leptons (electrons or muons) is presented, focusing on the challenging scenario in which the difference in mass between the two least massive neutralinos is approximately equal to the mass of the Z boson. The results are interpreted in simplified models of chargino-neutralino or neutralino pair production. For chargino-neutralino production, in the case when the lightest neutralino is massless, the combination yields an observed (expected) limit at the 95\% confidence level on the chargino mass of up to 650\,(570)\GeV, improving upon the individual analysis limits by up to 40\GeV. If the mass difference between the two least massive neutralinos is approximately equal to the mass of the Z boson in the chargino-neutralino model, the targeted search requiring three or more leptons obtains observed and expected exclusion limits of around 225\GeV on the second neutralino mass and 125\GeV on the lightest neutralino mass, improving the observed limit by about 60\GeV in both masses compared to the previous CMS result. In the neutralino pair production model, the combined observed (expected) exclusion limit on the neutralino mass extends up to 650--750\,(550--750)\GeV, depending on the branching fraction assumed. This extends the observed exclusion achieved in the individual analyses by up to 200\GeV. The combined result additionally excludes some intermediate gaps in the mass coverage of the individual analyses.
}

\hypersetup{%
pdfauthor={CMS Collaboration},%
pdftitle={Combined search for electroweak production of charginos and neutralinos in proton-proton collisions at sqrt(s) = 13 TeV},%
pdfsubject={CMS},%
pdfkeywords={CMS, physics, SUSY, electroweak}}

\maketitle

\section{Introduction}

Supersymmetry (SUSY)~\cite{Ramond:1971gb,Golfand:1971iw,Neveu:1971rx,Volkov:1972jx,Wess:1973kz,Wess:1974tw,Fayet:1974pd,Nilles:1983ge}
is an extension of the standard model (SM) of particle physics.
It posits a new symmetry such that for each boson (fermion) in the SM, there exists a fermionic (bosonic) superpartner.
Supersymmetry can potentially address several of the open questions in particle physics,
including the hierarchy problem~\cite{Dimopoulos:1995mi,Barbieri:2009ev,Papucci:2011wy}
and the unification of the gauge couplings at high energy scales~\cite{Buras:1977yy,Haber:1984rc}.
If $R$-parity~\cite{Farrar:1978xj} is conserved, the lightest SUSY particle (LSP) is stable
and could be a potential dark matter candidate~\cite{Goldberg:1983nd,Ellis:1983ew}.

This paper focuses on searches for electroweak production of SUSY particles,
under the assumption that the strongly-coupled SUSY particles are too massive to be directly produced.
The superpartners of the bosons from the SM SU(2) and U(1) gauge fields before electroweak symmetry breaking
are denoted as the winos and bino, respectively.
We consider SUSY models assuming two complex Higgs doublets, and the superpartners of the Higgs bosons are denoted as higgsinos.
The bino, winos, and higgsinos form mass eigenstates of two charginos (\chargino) and four neutralinos (\neutralino)
and in general can mix among one another.
In this paper, we focus on the lightest neutralino (\firstchi), the next-to-lightest neutralino (\secondchi),
and the lightest chargino (\firstcharg).
If the superpartners of the SM leptons, the sleptons, are much heavier than the charginos and neutralinos,
decays of the charginos and neutralinos proceed through the \PW, \PZ, and Higgs bosons.
The branching fractions of neutralinos to the \PZ{} and Higgs bosons depend
on the mixing among the bino, winos, and higgsinos to form mass eigenstates.

Searches performed at LEP exclude promptly-decaying charginos below a mass of 103.5\GeV~\cite{lepsusy}.
At the LHC, several searches have been performed by the
ATLAS~\cite{atlassos,Aad:2015eda,Aad:2015hea,Aad:2015qfa,Aad:2015jqa,ATLAS:2014fka,Aad:2014gfa,Aad:2014yka,Aad:2014iza,Aad:2014vma,Aad:2014nua,Aad:2013yna}
and CMS~\cite{Khachatryan:2014qwa,Khachatryan:2014mma,Khachatryan:2016hns,Khachatryan:2015pot,CMS:2015loa,Khachatryan:2016trj,Chatrchyan2013,Chatrchyan2015,ewkino2016,sos,razorhgg,hh4b,onz,wh1l}
Collaborations looking for direct production of charginos and neutralinos.
Given the various possible decay modes, a SUSY signal could simultaneously populate multiple final states.
This paper implements a statistical combination of the searches performed by CMS in Refs.~\cite{ewkino2016,sos,razorhgg,hh4b,onz,wh1l}
covering several final states
to improve upon the sensitivity of the individual analyses, particularly in models where the neutralino
has a nonzero branching fraction to both \PZ{} and Higgs bosons.
In addition, we present an extension of a search selecting events with three or more charged leptons~\cite{ewkino2016}.
It targets the difficult region of phase space where the difference in mass
between the \secondchi and \firstchi is approximately equal to the \PZ{} boson mass,
and the signal has similar kinematic properties to the dominant background of SM \wz production.
All searches use a data sample of LHC proton-proton collisions at $\sqrt{s} = 13\TeV$
collected by the CMS experiment in 2016, corresponding to an integrated luminosity of \fullLumi.
\section{Signal models}
\label{sec:sigs}

Simplified models of SUSY~\cite{bib-sms-2,bib-sms-3,bib-sms-4,Chatrchyan:2013sza}
are used to interpret the combined search results presented below.
In this paper, ``\PH'' refers to the 125\GeV scalar boson~\cite{Khachatryan:2016vau},
interpreted as the lightest CP-even state of an extended Higgs sector.
The \PH{} boson is expected to have SM-like properties if all of the other Higgs bosons are much heavier~\cite{Martin:1997ns}.
All signal models considered involve the production of two bosons (\PW, \PZ, or \PH) through SUSY decays, 
and we denote each model by the specific bosons produced.
The \PW, \PZ, and \PH{} bosons are always assumed to decay according to their SM branching fractions.
The sleptons are always assumed to have much higher masses than the charginos and neutralinos such that they do not contribute to the interactions.

\begin{figure}[htbp]
\centering
\includegraphics[width=0.4\textwidth]{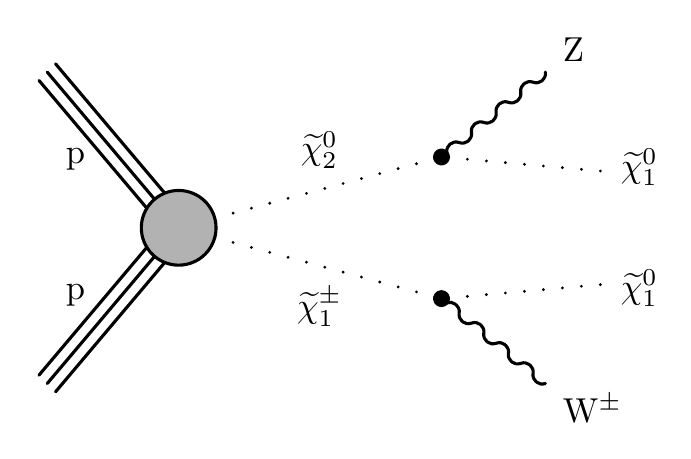}
\includegraphics[width=0.4\textwidth]{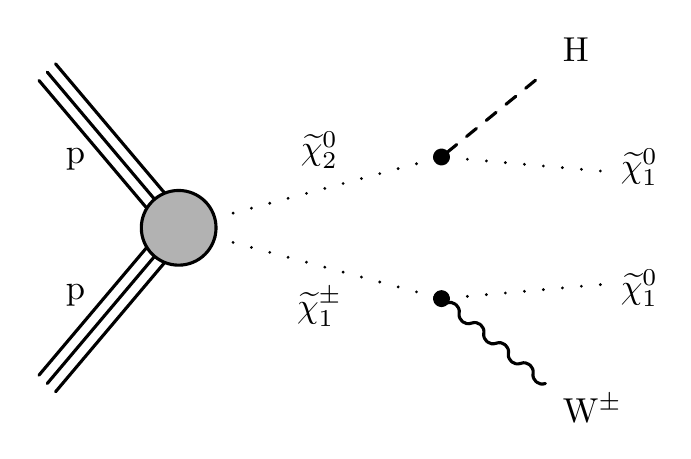}
\caption{Production of $\firstcharg\secondchi$ with the \firstcharg decaying to a \PW{} boson and the LSP, \firstchi, and the \secondchi decaying to
  either (left) a \PZ{} boson and the \firstchi or (right) a \PH{} boson and the \firstchi. }
\label{fig:c1n2}
\end{figure}

The first class of models assumes $\firstcharg\secondchi$ production.
The \firstchi is assumed to be the LSP.
The \firstcharg always decays to the \PW{} boson and the \firstchi, while
the \secondchi can decay to either of the \PZ{} or \PH{} bosons plus the \firstchi.
We consider three choices for the \secondchi\ decay:
a branching fraction of 100\% to $\PZ\firstchi$ (\wz topology),
of 100\% to $\PH\firstchi$ (\wh topology),
and of 50\% to each of these two decays (mixed topology).
This model is depicted in Fig.~\ref{fig:c1n2}, showing the two possible decays.
The production cross sections are computed in the limit of mass-degenerate winos \firstcharg and \secondchi,
and light bino \firstchi, with all other sparticles assumed to be heavy and decoupled.

\begin{figure}[htbp]
\centering
\includegraphics[width=0.3\textwidth]{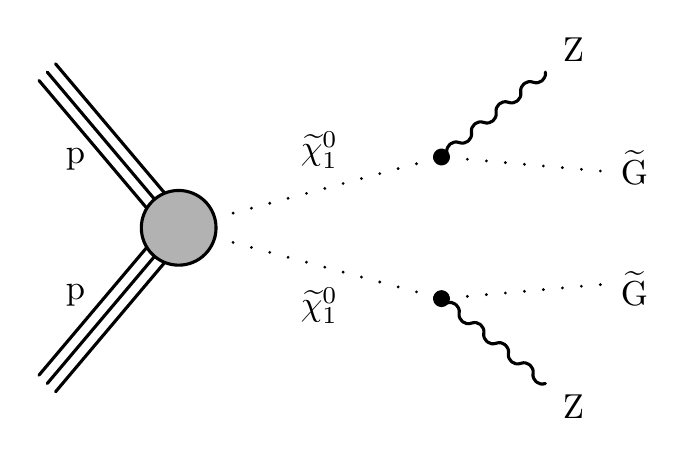}
\includegraphics[width=0.3\textwidth]{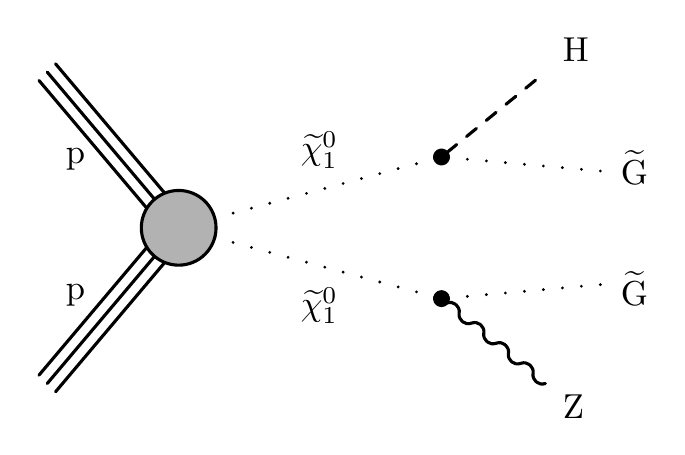}
\includegraphics[width=0.3\textwidth]{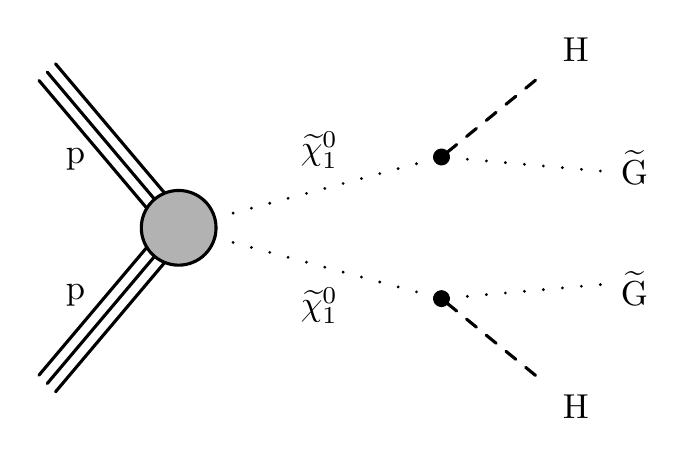}
\caption{A GMSB model with $\firstchi\firstchi$ pair production.  The two \firstchi particles decay into the \gravitino LSP and
(left) both to \PZ{} bosons, (center) a \PZ{} and a \PH{} boson, or (right) both to \PH{} bosons.}
\label{fig:n1n1}
\end{figure}

The second class of models assumes $\firstchi\firstchi$ production.
For bino- or wino-like neutralinos, the neutralino pair production cross section is very small, and thus
we consider a specific gauge-mediated SUSY breaking (GMSB) model with
quasidegenerate higgsinos as next-to-lightest SUSY particles
and an effectively massless gravitino (\gravitino) as the LSP~\cite{Matchev:1999ft,Meade:2009qv,Ruderman}.
In the production of any two of these, \firstcharg or \secondchi
decays immediately to \firstchi and low-momentum particles that do not impact the analysis,
effectively yielding pair production of $\firstchi\firstchi$.
The \firstchi then decays to a \gravitino and either a \PZ{} or \PH{} boson,
and we consider varying branching fractions from 100\% decay into the \PZ{} boson to 100\% decay into the \PH{} boson including intermediate values.
The possible decays in this model are shown in Fig.~\ref{fig:n1n1}.

The production cross sections for the GMSB scenario are computed
in a limit of mass-degenerate higgsino states \firstcharg, \secondchi, and \firstchi, with all the other sparticles assumed to be heavy and decoupled.
Following the convention of real mixing matrices and signed neutralino masses~\cite{Skands:2003cj}, we set the sign of the mass of \firstchi\ (\secondchi) to $+1$ ($-1$).
The lightest two neutralino states are defined as symmetric (antisymmetric) combinations of higgsino states by setting the product of the elements $N_{i3}$ and $N_{i4}$ of the neutralino mixing matrix $N$ to $+0.5$ ($-0.5$) for $i = 1$ (2).
The elements $U_{12}$ and $V_{12}$ of the chargino mixing matrices $U$ and $V$ are set to 1.

Cross section calculations to next-to-leading order (NLO)
plus next-to-leading-logarithmic (NLL) accuracy~\cite{Beenakker:1999xh,Fuks:2012qx,Fuks:2013vua,Bozzi:2007tea,Bozzi:2007qr,Bozzi:2006fw}
in perturbative quantum chromodynamics (QCD)
are used to normalize the signal samples for the results presented in Sections~\ref{sec:newmultilep} and~\ref{sec:interp}.
In this section, we present cross sections calculated to NLO accuracy~\cite{Fuks:2013vua} to demonstrate the dependence of the cross section values
on assumptions made in decoupling other SUSY particles.  The same qualitative conclusions also hold for the NLO+NLL calculations used
in the final results.

\begin{figure}[htbp]
\centering
\includegraphics[width=0.8\textwidth]{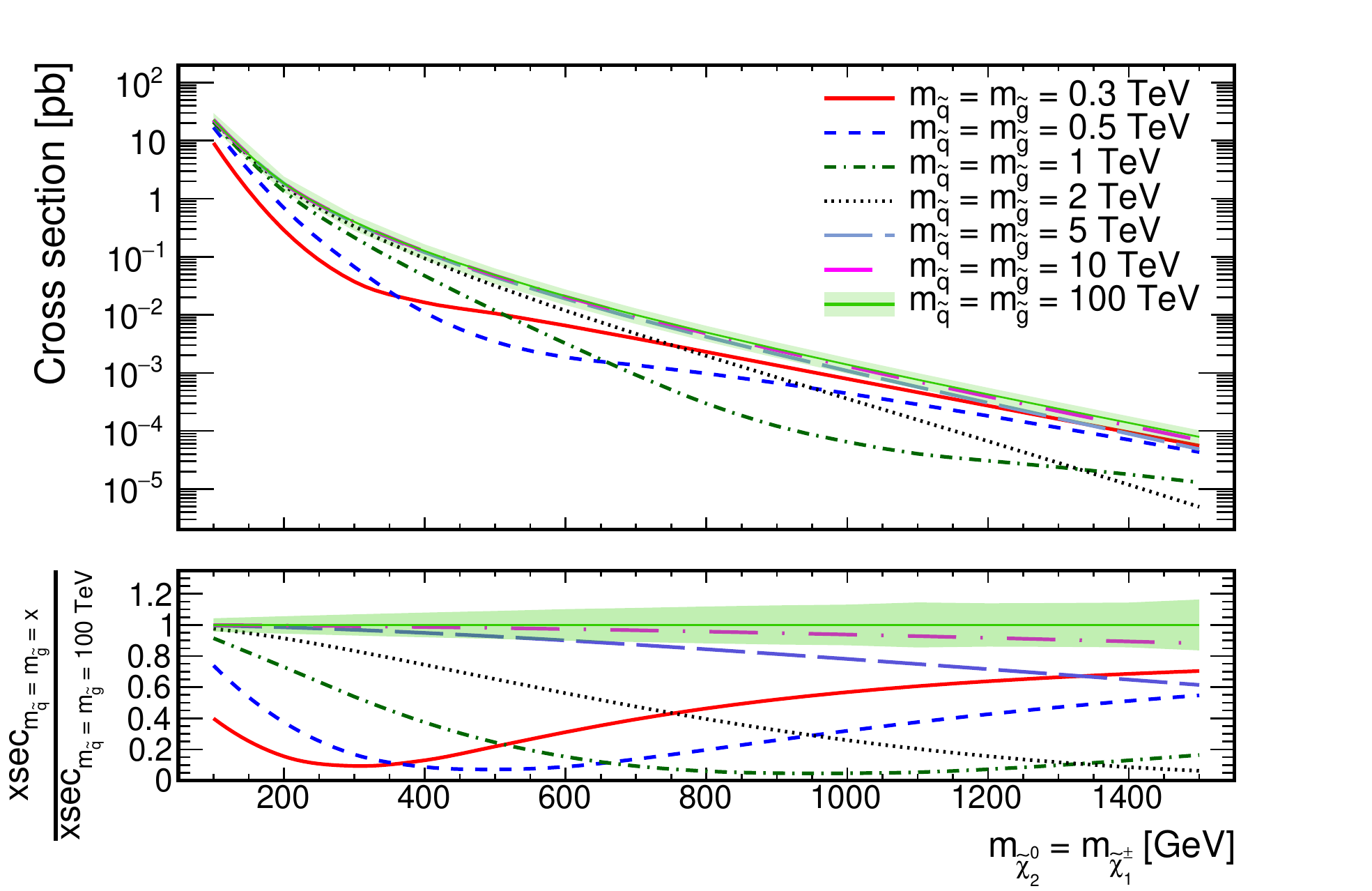}
\caption{Cross section for $\firstcharg\secondchi$ production at $\sqrt{s} = 13\TeV$ versus the wino mass,
  calculated to NLO accuracy in QCD with \textsc{Resummino}~\cite{Fuks:2013vua}.
  The \firstcharg and \secondchi are assumed to be mass-degenerate winos.
  The various curves show different assumptions on the masses of the squarks and gluinos, as described in the legend.
  The green band shows the theoretical uncertainty in the cross section calculation,
  from the variation of renormalization and factorization scales as well as parton density functions,
  for the 100\TeV squark and gluino mass assumption.
}
\label{fig:xsecs_strongmasses}
\end{figure}

Figure~\ref{fig:xsecs_strongmasses} shows the NLO cross section for $\firstcharg\secondchi$ production at $\sqrt{s} = 13\TeV$ assuming
mass-degenerate winos \firstcharg and \secondchi.
The various curves show different assumptions on the masses of squarks (\PSQ) and gluinos (\PSg), as described in the legend.
The cross section depends significantly on the masses of the strongly coupled particles until they reach masses of at least 10\TeV. For the range of \firstcharg and \secondchi masses considered here, the reduction can make up to 90\% in the cross section value.
This is due to large destructive interference effects from $t$-channel diagrams involving squark exchange.
The cross section calculation used in the interpretations of the analysis results assumes a mass of 100\TeV for the squarks and gluinos
to have them fully decoupled.
The obtained results would be less stringent if lower masses were assumed for the squarks and gluinos.
We performed the same study for $\firstcharg\secondchi$, $\firstcharg\firstchi$,
\firstchargpair, and $\secondchi\firstchi$ production with the assumption of mass-degenerate higgsinos \firstcharg, \secondchi, and \firstchi.
The dependence of the production cross section on the decoupling mass assumption was found to be much smaller in the higgsino case,
at most a few percent, and it is small compared to the uncertainty in the cross section calculation.

\section{The CMS detector}
\label{sec:cmsdetector}

The central feature of the CMS apparatus is a superconducting solenoid, 13\unit{m} in length and 6\unit{m} in diameter, that provides
an axial magnetic field of 3.8\unit{T}. The bore of the solenoid is outfitted with various particle detection systems. Charged-particle
trajectories are measured by silicon pixel and strip trackers, covering $0 < \phi < 2\pi$ in azimuth and $\abs{\eta} < 2.5$,
where the pseudorapidity $\eta$ is defined as $-\log [\tan(\theta/2)]$, with $\theta$ being the polar angle of the
trajectory of the particle with respect to the clockwise beam direction. A crystal electromagnetic calorimeter (ECAL) and a brass and scintillator
hadron calorimeter (HCAL) surround the tracking volume. The calorimeters provide energy
and direction measurements of electrons, photons, and hadronic jets. Muons are measured in gas-ionization detectors embedded in
the steel flux-return yoke outside the solenoid. The detector is nearly hermetic, allowing for energy balance measurements in the
plane transverse to the clockwise beam direction. A two-tier trigger system selects the most interesting $\Pp\Pp$ collision events for use
in physics analysis.
A more detailed description of the CMS detector, together with a definition of the coordinate system used and the relevant kinematic variables,
can be found in Ref.~\cite{Chatrchyan:2008zzk}.
\section{Event reconstruction and Monte Carlo simulation}
\label{sec:recosamples}

Event reconstruction is based on the particle-flow (PF) algorithm~\cite{Sirunyan:2017ulk},
which optimally combines information from the tracker, calorimeters, and muon systems to reconstruct and identify PF candidates,
\ie, charged and neutral hadrons, photons, electrons, and muons.
To select collision events, we require at least one reconstructed vertex.
The reconstructed vertex with the largest value of summed physics-object
$\pt^2$ is taken to be the primary $\Pp\Pp$ interaction vertex, where \pt is the transverse momentum with respect to the beam axis.
The physics objects are the objects returned by a jet finding
algorithm~\cite{antikt,FastJet} applied to all charged tracks
associated with the vertex, plus the corresponding associated missing transverse momentum.
The missing transverse momentum vector, \ptvecmiss, is defined as the negative vector sum of the momenta of all
reconstructed PF candidates projected onto the plane perpendicular to the proton beams. Its magnitude is referred to as \ptmiss.
Events with possible contributions from beam halo processes
or anomalous noise in the calorimeters can have large values of \ptmiss and are rejected using dedicated filters~\cite{Khachatryan:2014gga}.

Electron candidates are reconstructed starting from a cluster of energy deposits in the
ECAL. The cluster is then matched to a reconstructed track. The electron selection is based on the shower shape,
the ratio of energy measured in the HCAL to that measured in the ECAL, track-cluster matching,
and consistency between the cluster energy and the track momentum~\cite{Khachatryan:2015hwa}. Muon candidates are
reconstructed by performing a global fit that requires consistent hit patterns in the tracker and the muon system~\cite{MUOART}.
Photon candidates are reconstructed from a cluster of energy deposits in the ECAL, and they are required to pass
criteria based on the shower shape and the ratio of energy measured in the HCAL to that measured in the ECAL~\cite{Khachatryan:2015hwa}.
Hadronically decaying tau lepton candidates (\tauh) are reconstructed from PF candidates
with the ``hadron-plus-strips'' algorithm~\cite{Khachatryan:2015dfa}.
Electron, muon, photon, and \tauh candidates are required to be isolated from other particles,
and electron, muon, and \tauh candidates must satisfy requirements on the transverse and longitudinal impact parameters
relative to the primary vertex.

PF candidates are clustered to form jets using the anti-\kt clustering algorithm~\cite{antikt} with a distance parameter of 0.4,
as implemented in the {\FASTJET} package~\cite{FastJet}.
Identification of jets originating from b quarks (b jets) is performed with either the
combined secondary vertex (CSVv2) algorithm~\cite{BTV-16-002} or the DeepCSV algorithm~\cite{Guest:2016iqz}.
Data events are selected using a variety of triggers requiring the presence of electrons, muons, photons,
jets, or \ptmiss, depending on the final state targeted in each analysis.

Monte Carlo (MC) simulated samples are used in the various searches to estimate the
background from some SM processes, to assess systematic uncertainties in prediction methods that rely on data,
and to calculate the selection efficiency for signal models.
Most SM background samples are produced with the \MGvATNLO~v2.2.2 or~v2.3.3
generator~\cite{Alwall:2014hca}
at leading order (LO) or NLO accuracy in perturbative QCD,
including up to four additional partons in the matrix element calculations, depending on the process and calculation order.
Other samples are produced with the \POWHEG~v2~\cite{Melia:2011tj,Nason:2013ydw} generator without
additional partons in the matrix element calculations.
Standard model \wz production in particular is modeled with \MGvATNLO~v2.2.2 at NLO precision
for the search described in Section~\ref{sec:newmultilep}, which requires a precise description of initial-state radiation (ISR). 
In other cases, \POWHEG~v2 is used.
The NNPDF3.0 LO or NLO~\cite{Ball:2014uwa} parton distribution functions (PDFs) are used in the event generation.
Parton showering and fragmentation in all of these samples are performed using the \PYTHIA\ v8.212~\cite{Sjostrand:2007gs} generator
and the CUETP8M1 tune~\cite{Khachatryan:2015pea}.
A double counting of the partons generated with \MGvATNLO\ and those with \PYTHIA\ is removed
using the MLM~\cite{Alwall:2007fs} and the \textsc{FxFx}~\cite{Frederix:2012ps} matching schemes, in the LO and NLO samples, respectively.
Cross section calculations at NLO or next-to-NLO~\cite{Alwall:2014hca,Alioli:2009jeNew,Re:2010bp,Gavin:2010az,Gavin:2012sy,Czakon:2011xx}
are used to normalize the simulated background samples.

Signal samples are generated with \MGvATNLO\ at LO precision, including up to two additional partons
in the matrix element calculations.
Cross section calculations to NLO plus NLL accuracy~\cite{Borschensky:2014cia,Fuks:2012qx,Fuks:2013vua}
are used to normalize the signal samples.
For these samples
we improve on the modeling of ISR, which affects the total transverse
momentum of the system of SUSY particles ($\pt^\mathrm{ISR}$), by
reweighting the
 $\pt^\mathrm{ISR}$ distribution in these events.
This reweighting procedure is based on experimental studies of the \pt
of \PZ{} bosons~\cite{Chatrchyan:2013xna}.
The reweighting factors range between 1.18 (at $\pt^\mathrm{ISR} = 125\GeV$)
 and 0.78 (for $\pt^\mathrm{ISR} > 600\GeV$).
We take the deviation from 1.0 as the systematic uncertainty in the reweighting
procedure.

For both signal and background events, additional simultaneous proton-proton interactions (pileup)
are generated with \PYTHIA and superimposed on the hard collisions.
The response of the CMS detector for SM background samples is simulated using a \GEANTfour-based model~\cite{Geant}, while that for new
physics signals is performed using the CMS fast simulation package~\cite{fastsim}.
All simulated events are processed with the same chain of reconstruction programs as used for collision data.
Corrections are applied to simulated samples to account for differences between
the trigger, b tagging, and lepton and photon selection efficiencies measured in data and the \GEANTfour simulation.
Additional differences arising from the fast simulation modeling of selection efficiencies, as well as
from the modeling of \ptmiss, are corrected in the fast simulation and included in the systematic uncertainties considered.
\section{Individual searches}
\label{sec:searches}

The experimental searches included in the combination are briefly described here.
Table~\ref{tab:searches_models} lists which searches are used to place exclusion limits
for each of the topologies introduced in Section~\ref{sec:sigs}.
The selections for all searches were checked to be mutually exclusive,
such that no events fulfill the signal region requirements for more than one search.
No significant deviations from the SM predictions were observed in any of these searches.

\begin{table}[htb]
\centering
  \topcaption{\label{tab:searches_models} Summary of all experimental searches considered in the combination (rows),
    and the signal topologies for which each search is used in the combined results (columns).
    The searches are described in Sections~\ref{sec:lepbb} through~\ref{sec:hgg} and Section~\ref{sec:newmultilep}.
    The ${\geq}3\ell$ search described in Section~\ref{sec:oldmultilep} is used for all signal topologies except
    for \wz, where the reoptimized search strategy from Section~\ref{sec:newmultilep} is employed instead.
  }
\begin{tabular}{l|ccccc}
   & \multicolumn{5}{c}{Signal topology} \\
  Search & \wz & \wh & \zz & \zh & \hh \\
  \hline
  $1\ell$ 2\PQb &  & \checkmark &  &  &  \\
  4\PQb &  &  &  &  & \checkmark \\
  $2\ell$ on-\PZ & \checkmark &  & \checkmark & \checkmark &  \\
  $2\ell$ soft & \checkmark &  &  &  &  \\
  ${\geq}3\ell$ & \checkmark & \checkmark & \checkmark & \checkmark & \checkmark \\
  \hgg &  & \checkmark &  & \checkmark & \checkmark \\
\end{tabular}
\end{table}

\subsection{Search for one lepton, two b jets, and \texorpdfstring{\ptmiss}{missing transverse momentum}}
\label{sec:lepbb}

The ``$1\ell$ 2\PQb'' search~\cite{wh1l}, targeting the \wh\ topology,
selects events with exactly one charged lepton (\Pe{} or \Pgm), exactly two b jets, and large \ptmiss.
The invariant mass of the two b jets is required to be consistent with the mass of the \PH{} boson.
Kinematic variables are used to suppress backgrounds, which predominantly come from
dileptonic decays in \ttbar production.
Two exclusive signal regions are defined based on \ptmiss: $125 \leq \ptmiss < 200\GeV$
and $\ptmiss \geq 200\GeV$.
The SM backgrounds are predicted using MC simulation,
with the predictions validated in data control regions distinct from the signal region.

\subsection{Search for four b jets and \texorpdfstring{\ptmiss}{missing transverse momentum}}
\label{sec:fourb}

The ``4\PQb'' search~\cite{hh4b}, targeting the \hh\ topology,
selects events with exactly four or five jets, with at least two of them identified as b jets,
large \ptmiss, and no charged leptons.
In each event, the four jets with the highest b tagging discriminator scores are considered to form dijet \PH{} candidates.
There are three possible groupings to make two pairs of jets.
The grouping is selected to minimize the difference between the invariant masses of the two dijet pairs,
and the difference in masses is required to be less than 40\GeV.
The average invariant mass of the two pairs is then required to be consistent with the mass of the \PH{} boson.
Exclusive signal regions are defined based on the number of b jets (three or at least four)
and multiple bins in \ptmiss.
The primary background to this search comes from semileptonic decays in \ttbar production, with smaller contributions
from \PW{} or \PZ{} production in association with jets and from QCD multijet production.
The backgrounds are predicted using data control samples that require either exactly two b jets or
an average dijet invariant mass inconsistent with the \PH{} boson.

\subsection{Search for two leptons consistent with a \PZ{} boson, jets, and \texorpdfstring{\ptmiss}{missing transverse momentum}}
\label{sec:onz}

The ``$2\ell$ on-\PZ'' search~\cite{onz}, targeting the \wz, \zz, and \zh topologies,
selects events with exactly two opposite-sign, same-flavor (OSSF) leptons (\ElEl or \MuMu) consistent with the \PZ boson mass,
at least two jets, and large \ptmiss.
In the signal region targeting the \wz and \zz topologies, two jets are required to have an invariant mass 
less than 110\GeV to be compatible with the \PW\ and \PZ\ boson masses, and events with b jets are rejected.
To target the \zh topology, events are required to have two b jets with an invariant mass 
less than 150\GeV to be compatible with the \PH{} boson mass.
Signal regions are defined with multiple exclusive bins in \ptmiss.
The backgrounds fall into three categories. First, flavor symmetric backgrounds, such as \ttbar production, yield \EM events
at the same rate as \ElEl and \MuMu events combined, and they are predicted from a data control sample of \EM events.
Second, events with a \PZ{} boson and mismeasured jets give instrumental \ptmiss, and they are predicted from a data control sample
of \cPgg+jets events.  
Third, events with a \PZ{} boson and at least one prompt neutrino, 
arising from processes such as \wz, \zz, and \ttz\ production, are estimated using simulation.

\subsection{Search for two soft leptons and \texorpdfstring{\ptmiss}{missing transverse momentum}}
\label{sec:sos}

The ``$2\ell$ soft'' search~\cite{sos} selects events with exactly two low-\pt\ leptons (\ElEl or \MuMu in the relevant selections),
jets, and large \ptmiss.
It targets the \wz topology where the mass difference between \secondchi and \firstchi is small such that the \PW{} and \PZ{} bosons are off-shell, and the observable decay products have low momentum.
The leptons are required to satisfy $5 < \pt < 30\GeV$ and have an invariant mass
in the range $4 < \mll < 50\GeV$, strongly suppressing SM backgrounds while retaining good acceptance for compressed signal scenarios.
Additional kinematic requirements are applied to further reduce backgrounds, and the relevant signal regions are binned
in \mll\ and \ptmiss.
The largest backgrounds arise from $\PZ/\gamma^{*}$ and \ttbar production, as well as misidentification of nonprompt leptons.
The first two are predicted from simulation with constraints from data control regions,
while the latter is predicted entirely using data.

\subsection{Search for three or more leptons, and \texorpdfstring{\ptmiss}{missing transverse momentum}}
\label{sec:oldmultilep}

The ``${\geq}3\ell$'' search~\cite{ewkino2016} selects events with three or more leptons (\Pe, \Pgm, and up to two \tauh) and large \ptmiss.
Several exclusive categories are defined based on the number of leptons, lepton flavor and charge, the presence of an OSSF pair,
and kinematic variables such as the invariant mass of the OSSF pair and \ptmiss.
Events with a b jet are rejected to reduce the background from \ttbar production.
The various categories are designed to give this search sensitivity for a wide range of new physics models,
including all of the topologies introduced in Section~\ref{sec:sigs}.
The best performance is seen in the \wz and \zz models, while the lower branching fraction of the \PH{} boson
to leptons reduces the sensitivity to other models.
The SM backgrounds in this search vary across the categories, and the most important for the relevant regions
in these interpretations are SM \wz and \zz production, and events with misidentified nonprompt leptons.
The former are predicted using simulation, which in case of \wz is validated in a set of dedicated control regions,
while the latter are predicted entirely from data.

A further optimization of this analysis has been performed for the \wz topology in the case where
the difference in the masses of \secondchi and \firstchi is equal to the \PZ{} boson mass,
focusing on a category selecting events with three light-flavor leptons (\Pe, \Pgm).
This update is presented in Section~\ref{sec:newmultilep}.

\subsection{Search for a \PH{} boson decaying to diphotons and \texorpdfstring{\ptmiss}{missing transverse momentum}}
\label{sec:hgg}

The ``\hgg'' search~\cite{razorhgg} selects events with two photons consistent with the \PH{} boson mass, along with jets and large \ptmiss.
Events are categorized based on the \pt\ of the diphoton system, the expected resolution
on the diphoton mass, the presence of two b jets compatible with the \PH{} or \PZ{} boson masses,
and the razor kinematic variables~\cite{Chatrchyan:2012uea,Chatrchyan:2014goa}.
It exhibits sensitivity to the \wh, \zh, and \hh topologies.
The background arises either from $\gamma$+jets or SM \PH{} boson production.
The former is estimated using a fit to the diphoton mass spectrum in a wider range than the signal window,
while the latter is predicted using simulation.

\section{Search for three light leptons consistent with \wz production and \texorpdfstring{\ptmiss}{missing transverse momentum}}
\label{sec:newmultilep}

The multilepton search described in Section~\ref{sec:oldmultilep} contains a category selecting
events with three light-flavor leptons (\Pe, \Pgm), two of which must form an
OSSF pair. This final state aims to provide sensitivity
for a variety of SUSY models, including the \wz topology depicted
in Fig.~\ref{fig:c1n2} (left).
The dominant background in this search category is SM \wz production.

Exclusion limits on the \wz topology were placed in Ref.~\cite{ewkino2016}, and
the sensitivity was found to be significantly reduced for $\msecondchi - \mfirstchi \approx m_{\PZ}$,
referred to here as the ``\wz corridor.''
In this case, SUSY signal is kinematically similar to the SM background.
We present here a further optimization of the search for the \wz
topology designed to target this challenging region of phase space.
The search methodology remains the same as in Ref.~\cite{ewkino2016},
but the event categorization has been updated as described below.

We require events to have three light-flavor leptons with two forming an OSSF pair.
Events are categorized using the following kinematic variables:
\ptmiss, the invariant mass \mll\ of the OSSF pair, and the
transverse mass \MT\ of the third lepton computed with respect to \ptmiss.
Three bins in \mll are defined to separate contributions from on- and
off-shell \PZ{} boson decays,
and three bins are defined in \MT to separate the SM \PW{} boson contribution.

\begin{figure}[!hbtp]
\centering
\includegraphics[width=0.45\textwidth]{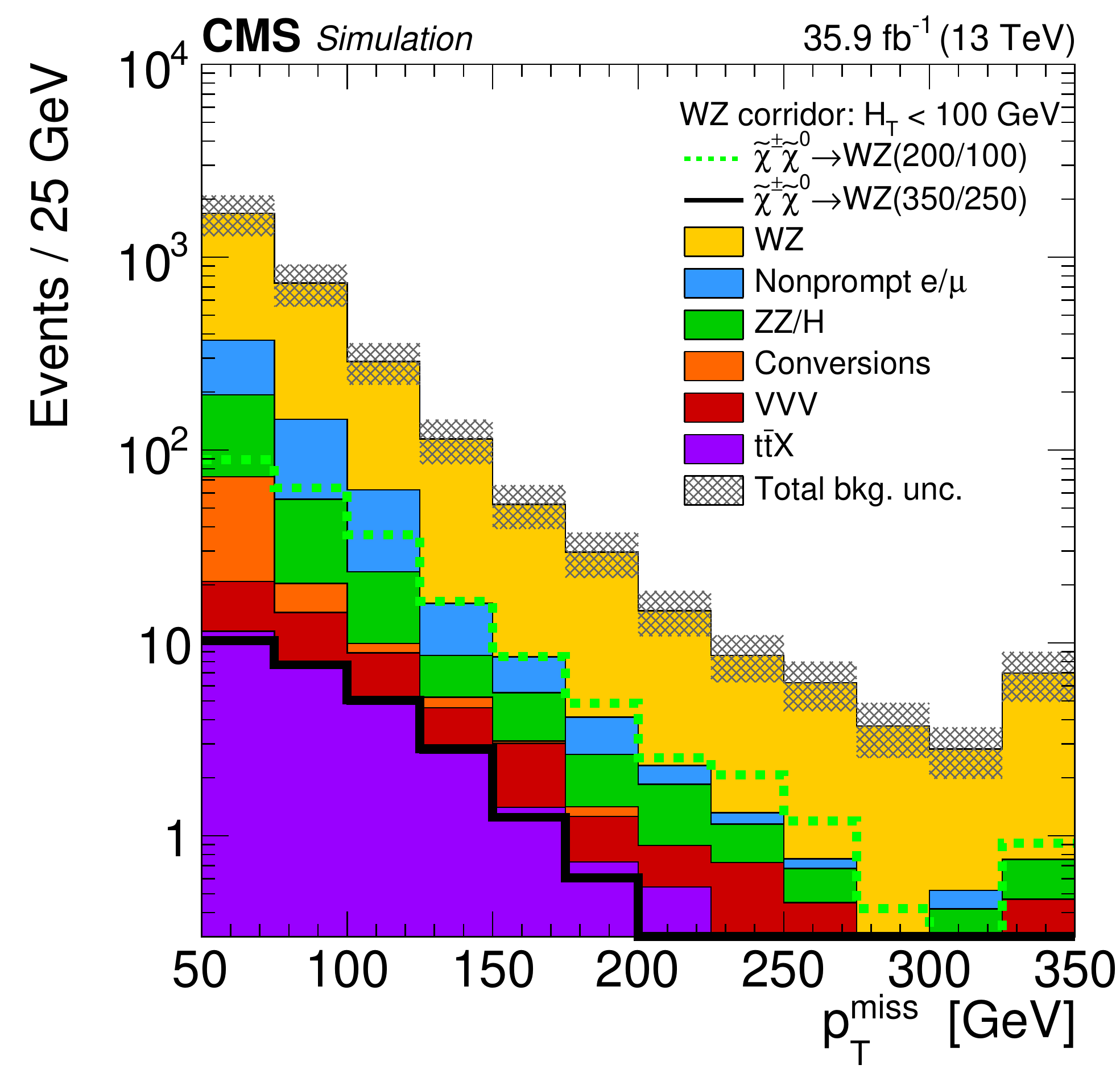}
\includegraphics[width=0.45\textwidth]{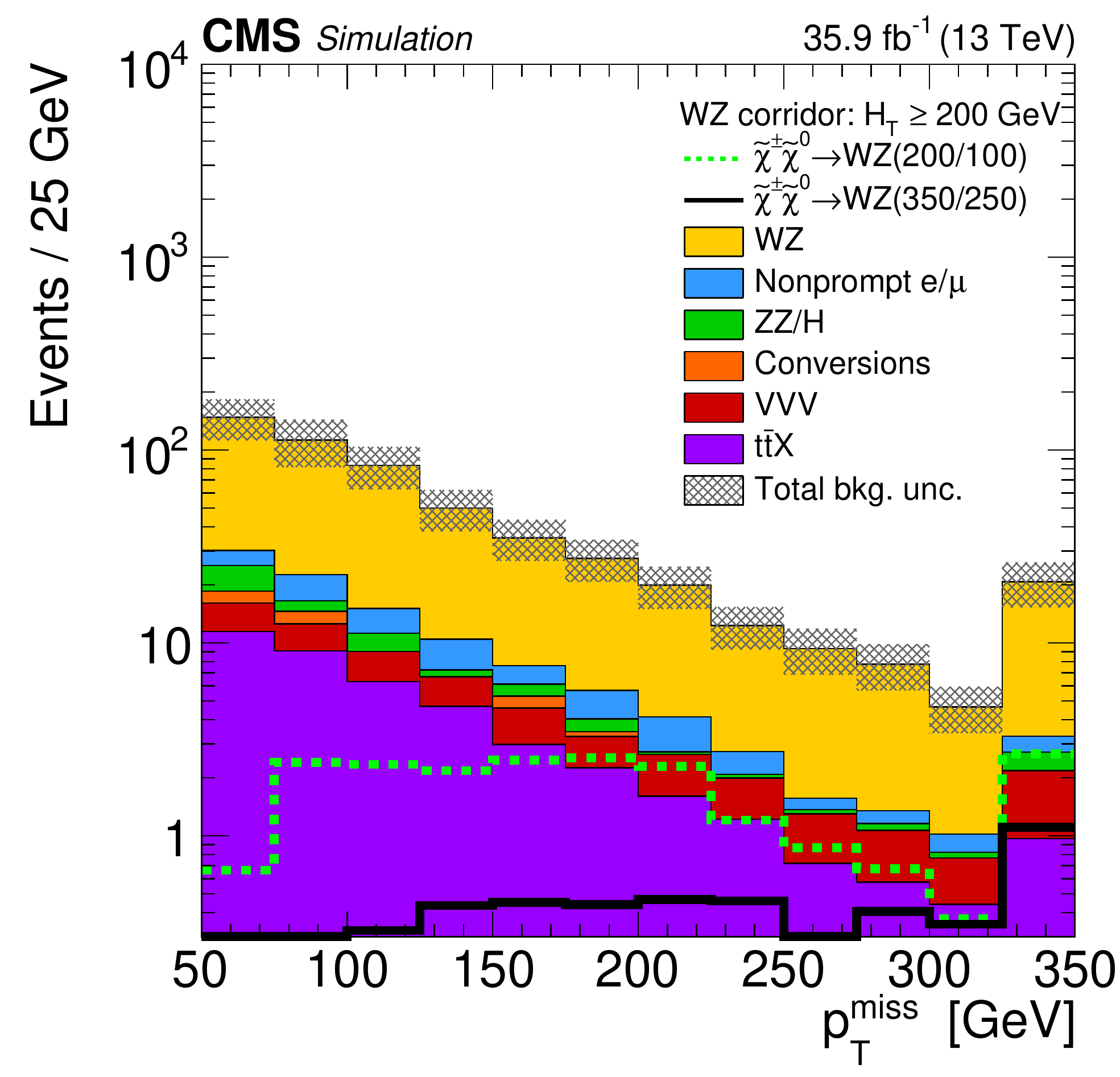}
\caption{Distributions of \ptmiss for two representative
signal points in the \wz corridor as well as the expected SM background for $\HT < 100$
(left) and $\geq 200\GeV$ (right). 
The mass values for the signal points are given as $(\msecondchi / \mfirstchi)$ in \GeV.
For larger values of \HT, the
shape difference between signal and background becomes more pronounced due to
the presence of \firstchi\ LSPs with large Lorentz boost.}
\label{fig:newmultilep:expkin}
\end{figure}

To improve the separation between signal and background in the \wz corridor, we exploit
ISR by further categorizing the events in \HT, the scalar \pt sum of the
jets with $\pt > 30\GeV$.
Due to the presence of the \firstchi LSPs, signal model points in the \wz corridor will tend to have more events
at high values of \ptmiss and \MT than the SM background for the same value of \HT,
with the effect becoming relevant at $\mfirstchi \approx m_{\PZ}$ and more pronounced at higher \HT.
This is demonstrated in Fig.~\ref{fig:newmultilep:expkin},
which shows the expected distributions of \ptmiss for background and two signal model points
after requiring (left) $\HT < 100\GeV$ and (right) $\geq 200\GeV$.
The \HT categorization is applied in the regions
$\mll < 75\GeV$ and $75 \leq \mll < 105\GeV$.
The full set of search regions is summarized in Table~\ref{tab:newmultilep:categories}.

\begin{table}[tbh]
\centering
\topcaption{Definition of the search regions (SRs) optimized for the \wz corridor
in the \wz signal topology. Events must have three leptons (\Pe, \Pgm) forming at least
one OSSF pair and they are categorized in \mll, \MT, \ptmiss and \HT.
Where ranges of values are given, the lower bound is inclusive while the upper bound is exclusive,
\eg, $75 \leq \mll < 105\GeV$.
}
\label{tab:newmultilep:categories}
\resizebox{0.8\textwidth}{!}{
\begin{tabular}{|c|c|c|c|c|c|}
\hline
$\mll$ (\GeVns{}) & $\MT$ (\GeVns{}) & $\ptmiss$ (\GeVns{}) & $\HT < 100\GeV$ & $100 \leq \HT < 200\GeV $ & $\HT\geq200\GeV $\\
\hline\hline
\multirow{11}{*}{0--75  }  & \multirow{4}{*}{0--100}   & 50--100  & \multicolumn{2}{c|}{SR 01} & \multirow{4}{*}{SR 12} \\ \cline{3-5}
                            &                               & 100--150 & \multicolumn{2}{c|}{SR 02} &  \\ \cline{3-5}
                            &                               & 150--200 & \multicolumn{2}{c|}{SR 03} &  \\ \cline{3-5}
                            &                               & $\geq$200 & \multicolumn{2}{c|}{SR 04} &  \\ \cline{2-6}
                            & \multirow{3}{*}{100--160} & 50--100  & \multicolumn{2}{c|}{SR 05} & \multirow{3}{*}{SR 13} \\ \cline{3-5}
                            &                               & 100--150 & \multicolumn{2}{c|}{SR 06} &  \\ \cline{3-5}
                            &                               & $\geq$150 & \multicolumn{2}{c|}{SR 07} &  \\ \cline{2-6}
                            & \multirow{4}{*}{$\geq$160}    & 50--100  & \multicolumn{2}{c|}{SR 08} & \multirow{4}{*}{SR 14} \\ \cline{3-5}
                            &                               & 100--150 & \multicolumn{2}{c|}{SR 09} &  \\ \cline{3-5}
                            &                               & 150--200 & \multicolumn{2}{c|}{SR 10} &  \\ \cline{3-5}
                            &                               & $\geq$200 & \multicolumn{2}{c|}{SR 11} &  \\ \hline
\multirow{18}{*}{75--105}  & \multirow{6}{*}{0--100}   & 50--100  & (WZ CR)                 & SR 27 & \multirow{2}{*}{SR 40} \\ \cline{3-5}
                            &                               & 100--150 & SR 15                  & SR 28 &        \\ \cline{3-6}
                            &                               & 150--200 & SR 16                  & SR 29 & \multirow{2}{*}{SR 41} \\ \cline{3-5}
                            &                               & 200--250 & SR 17                  & SR 30 &        \\ \cline{3-6}
                            &                               & 250--350 & \multirow{2}{*}{SR 18} & \multirow{2}{*}{SR 31} & SR 42 \\ \cline{3-3}\cline{6-6}
                            &                               & $\geq$350 &                         &        & SR 43 \\ \cline{2-6}
                            & \multirow{6}{*}{100--160} & 50--100  & SR 19                  & SR 32 & SR 44 \\ \cline{3-6}
                            &                               & 100--150 & SR 20                  & SR 33 & SR 45 \\ \cline{3-6}
                            &                               & 150--200 & SR 21                  & SR 34 & SR 46 \\ \cline{3-6}
                            &                               & 200--250 & \multirow{3}{*}{SR 22} & \multirow{3}{*}{SR 35} & SR 47 \\ \cline{3-3}\cline{6-6}
                            &                               & 250--300 &                         &       & SR 48 \\ \cline{3-3}\cline{6-6}
                            &                               & $\geq$300 &                         &       & SR 49 \\ \cline{2-6}
                            & \multirow{6}{*}{$\geq$160}    & 50--100  & SR 23                  & SR 36 & SR 50 \\ \cline{3-6}
                            &                               & 100--150 & SR 24                  & SR 37 & SR 51 \\ \cline{3-6}
                            &                               & 150--200 & SR 25                  & SR 38 & SR 52 \\ \cline{3-6}
                            &                               & 200--250 & \multirow{3}{*}{SR 26} & \multirow{3}{*}{SR 39} & SR 53 \\ \cline{3-3}\cline{6-6}
                            &                               & 250--300 &                         &        & SR 54 \\ \cline{3-3}\cline{6-6}
                            &                               & $\geq$300 &                         &        & SR 55 \\ \hline

\multirow{ 3}{*}{$\geq$105} & 0--100                    & $\geq$50  & \multicolumn{3}{c|}{SR 56} \\ \cline{2-6}
                            & 100--160                  & $\geq$50  & \multicolumn{3}{c|}{SR 57} \\ \cline{2-6}
                            & $\geq$160                     & $\geq$50  & \multicolumn{3}{c|}{SR 58} \\ \hline
\end{tabular}}
\end{table}

The dominant background in this search is SM \wz production, which provides a signature
very similar to the signal process in the form of three isolated leptons
and substantial \ptmiss due to the neutrino from the \PW{} boson decay. This background is
estimated from simulation, while two control regions are used to assess the overall normalization
and to validate the modeling of events at large values of \ptmiss, \MT, or both.
Further backgrounds arise from misidentification of nonprompt leptons from processes like \ttbar production, external and internal photon conversions,
and rare SM processes such as triboson production, \ttw, and \ttz. The contribution of the nonprompt
lepton background is predicted using the ``tight-to-loose'' ratio
method~\cite{Khachatryan:2016kod}, which relies entirely on data. External and internal photon conversions as well as rare SM processes
are predicted from simulation, and a dedicated data control region is used to constrain the normalization
of the conversion background.

The SM \wz background normalization is constrained in a data control region requiring
$75 \leq \mll < 105\GeV$, $\MT < 100\GeV$, $35 < \ptmiss < 100\GeV$, and
$\HT < 100\GeV$. The fraction of selected background events arising from SM \wz production
in this region is approximately 86\%. The validation of the \ptmiss and \MT\ shape
modeling is done using a data control sample enriched in $\PW\gamma$ events,
with the remainder of events coming mainly from $\WJ$ production.
A photon with $\pt > 40\GeV$ is required together with a lepton and $\ptmiss \geq 50\GeV$,
corresponding to a leptonic \PW{} boson decay. The minimum photon \pt\ threshold ensures
that the photon does not arise from final-state radiation. The
motivation behind this selection is that the \PW{} boson \MT\ distribution
in both $\PW\gamma$ and $\WJ$ events is found to be consistent with that of SM \wz production.
A systematic uncertainty is assigned to the signal region bins with high \MT\ and \ptmiss based on the statistical precision of this control region.

\begin{figure}[!hbtp]
\centering
\includegraphics[width=0.45\textwidth]{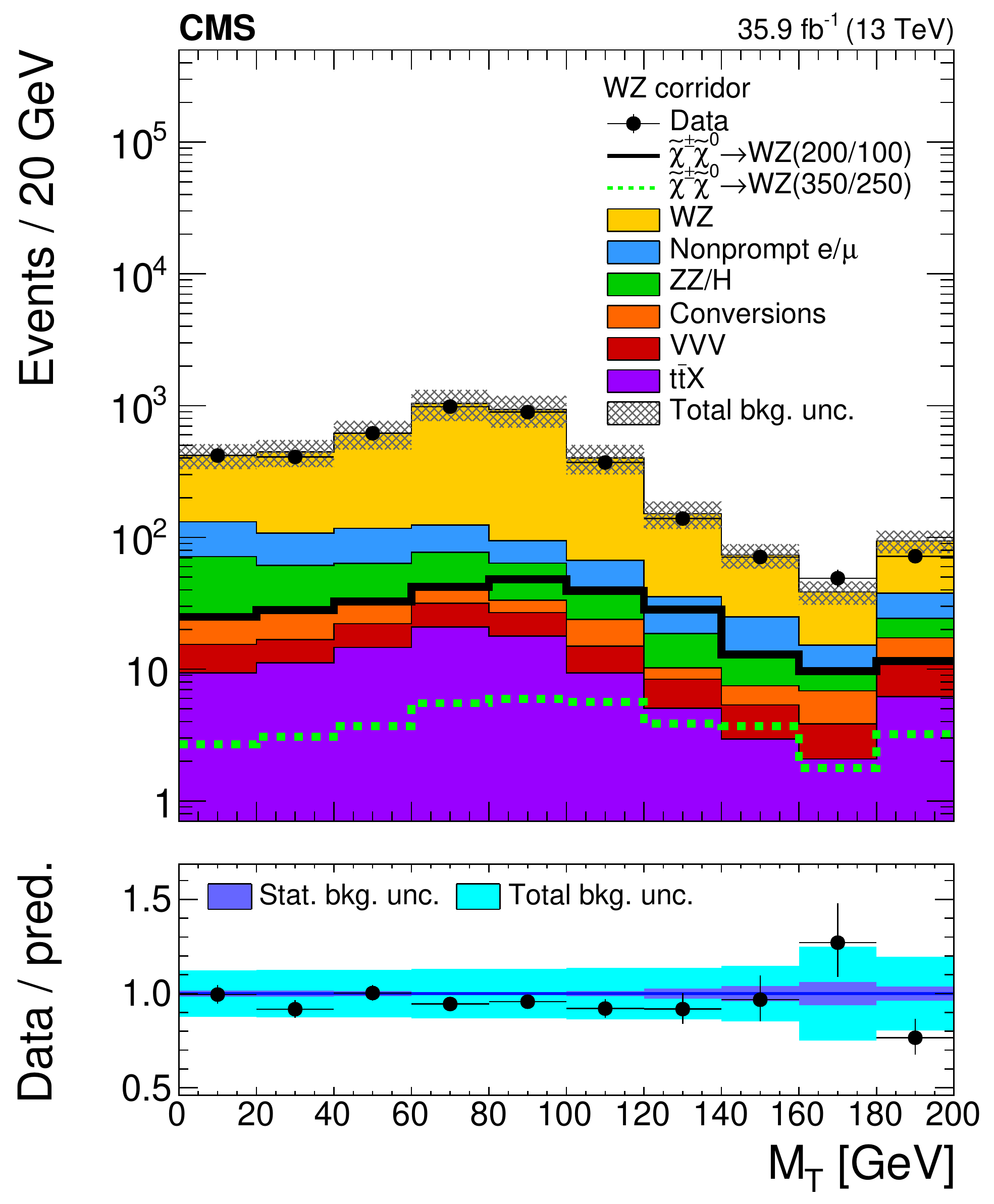}
\includegraphics[width=0.45\textwidth]{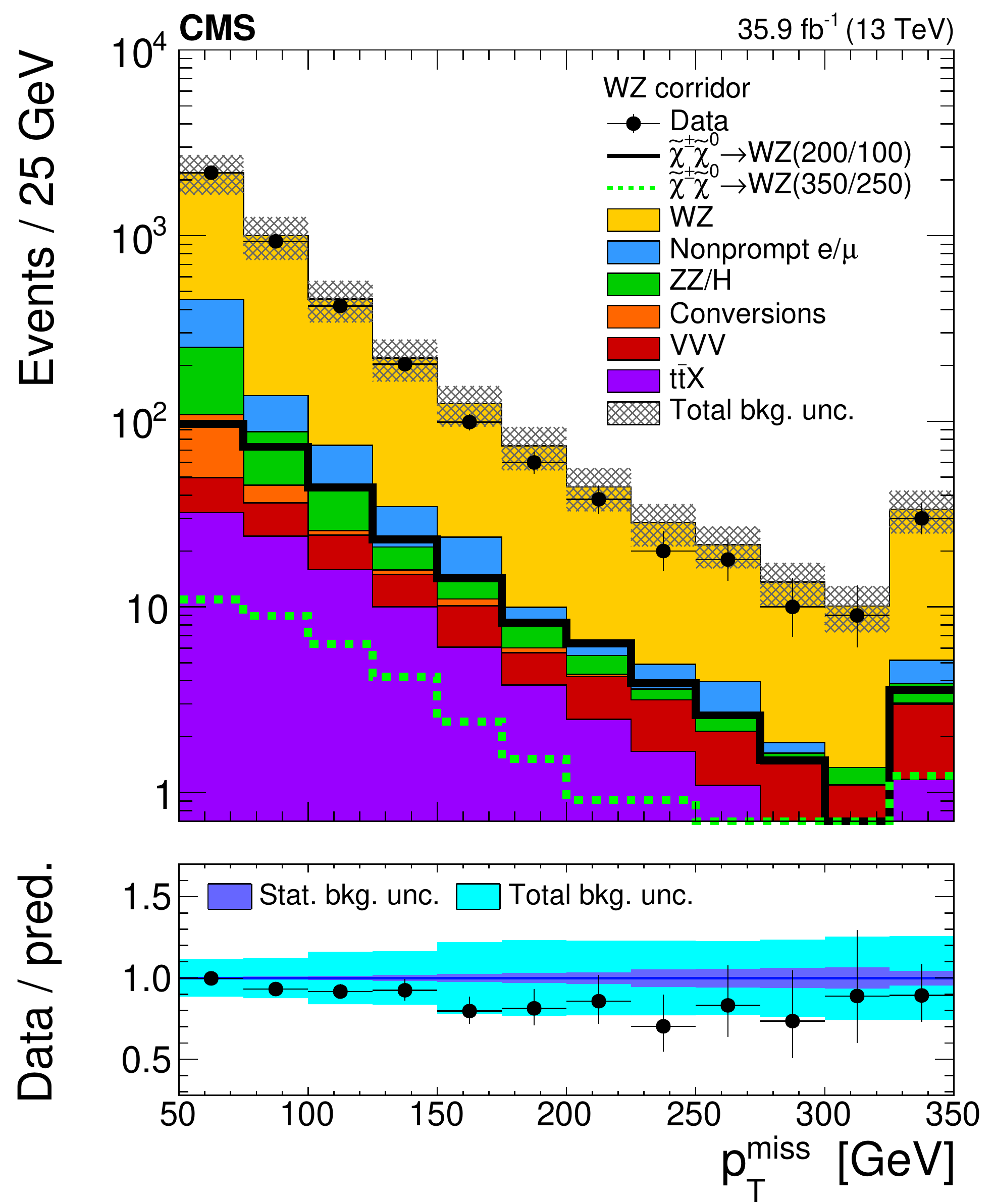} \\
\includegraphics[width=0.45\textwidth]{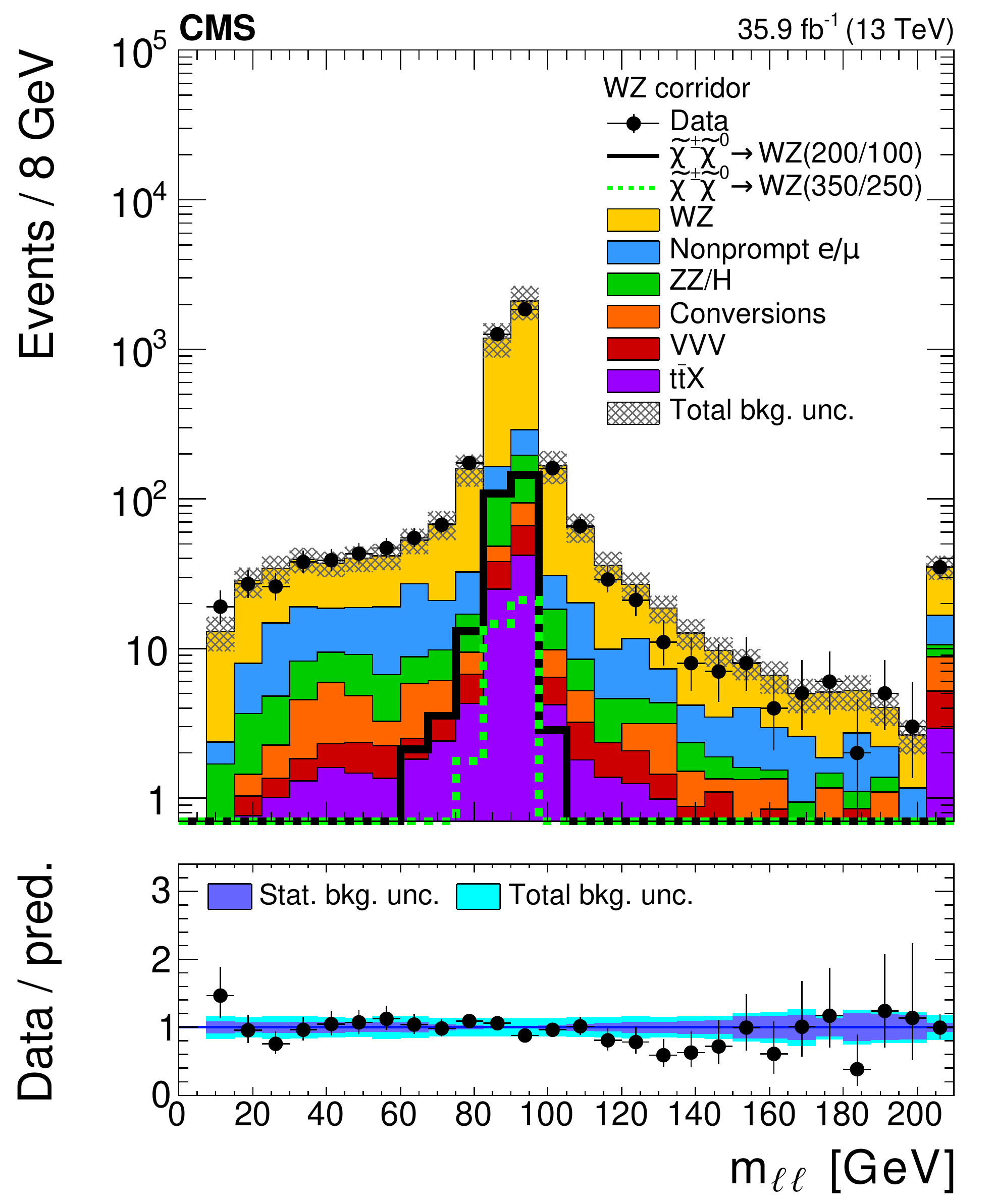}
\includegraphics[width=0.45\textwidth]{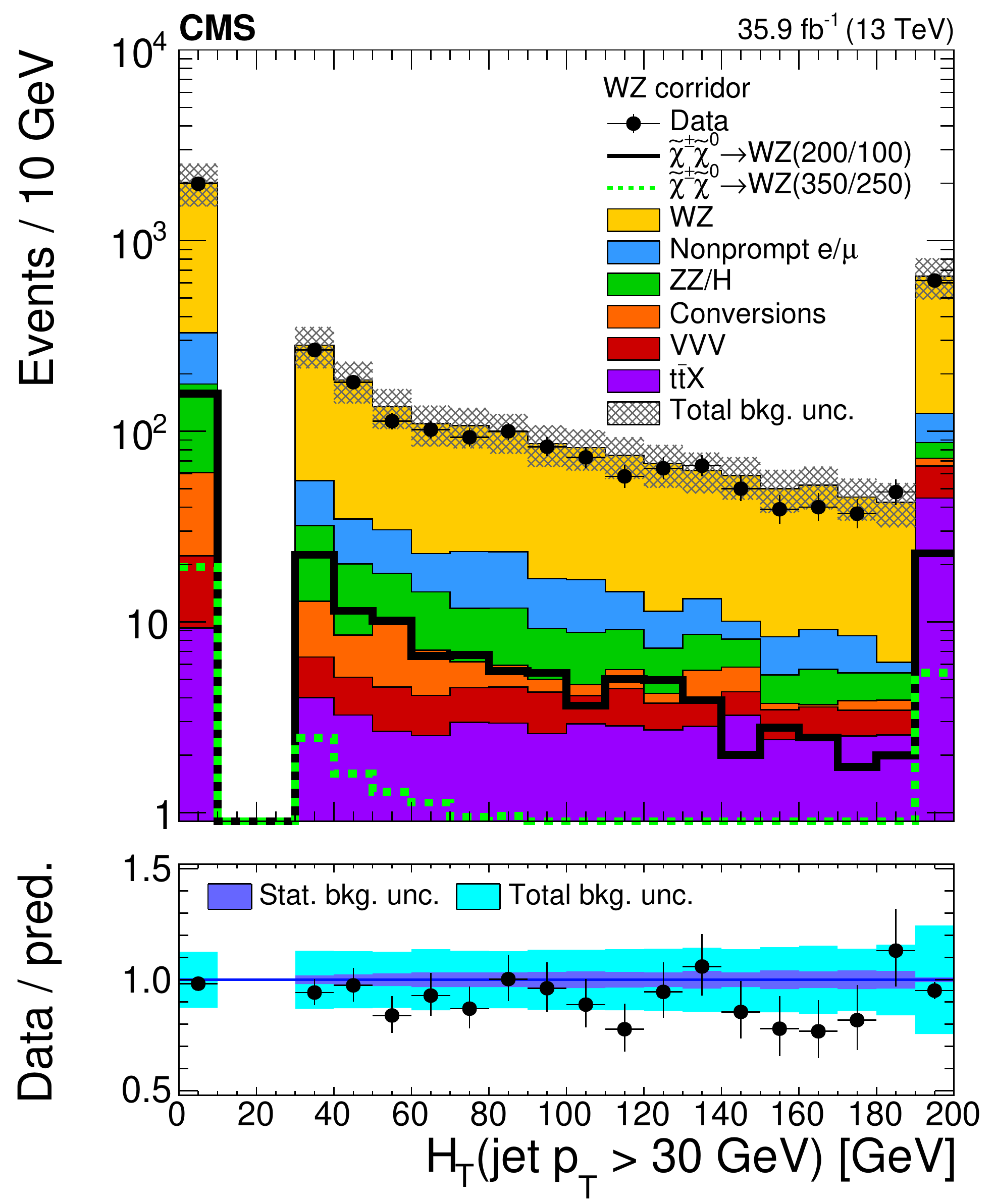}
\caption{Distributions of the transverse mass of the third lepton with respect to \ptmiss (upper left),
the \ptmiss (upper right), the \mll\ of the OSSF pair (lower left), and
the \HT\ (lower right). Distributions for two signal mass points in the \wz corridor
are overlaid for illustration.
The mass values for the signal points are given as $(\msecondchi / \mfirstchi)$ in \GeV.
The bottom panel shows the ratio of observed data to predicted yields.
The dark purple band shows the statistical uncertainty in the background prediction,
while the light blue band shows the total uncertainty.}
\label{fig:newmultilep:results:kin}
\end{figure}

\begin{figure}[!hbtp]
\centering
\includegraphics[width=0.95\textwidth]{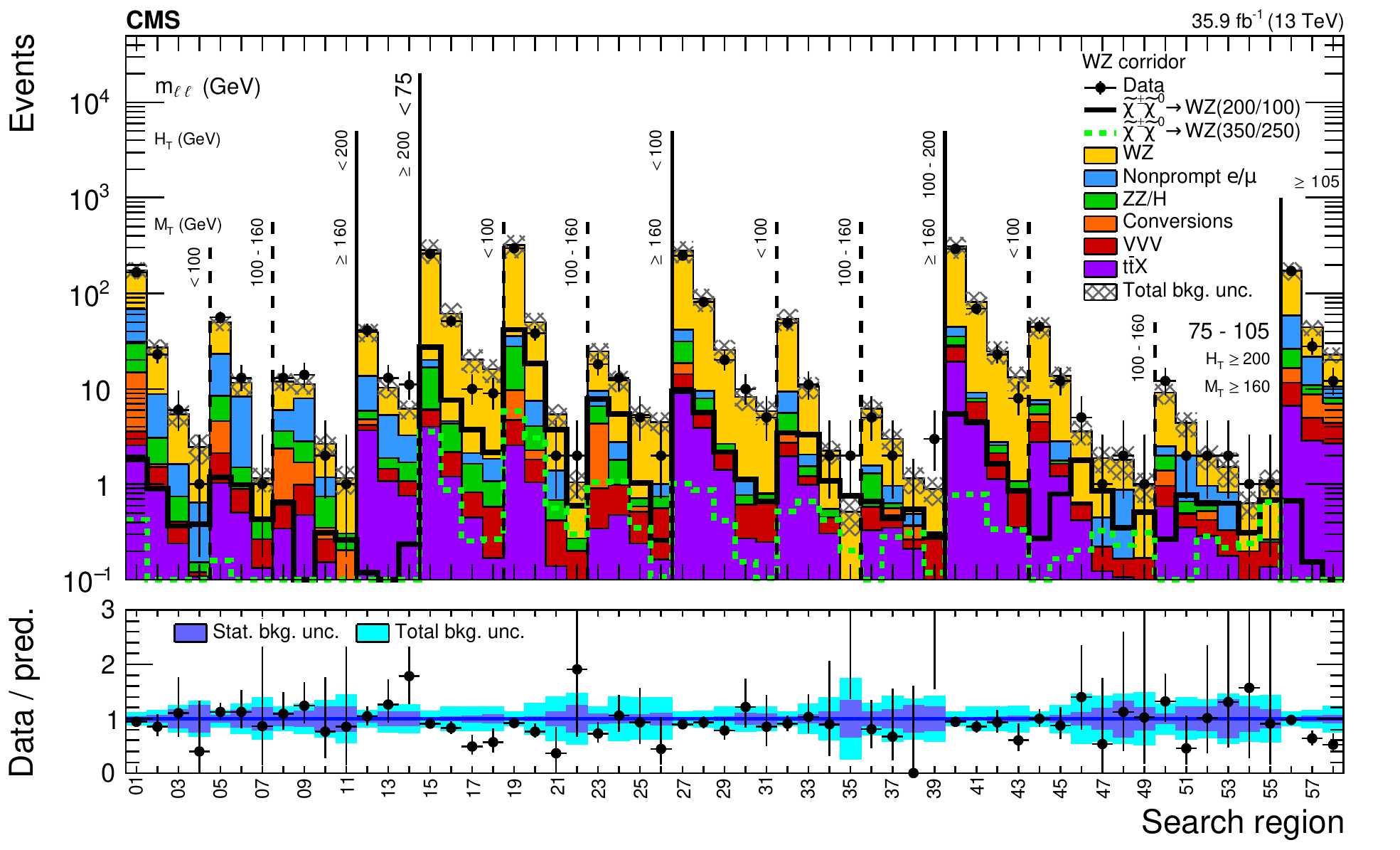}
\caption{Expected and observed yield comparison in the search regions. Two
example signal mass points along the \wz corridor are overlaid for illustration.
The mass values for the signal points are given as $(\msecondchi / \mfirstchi)$ in \GeV.
The bottom panel shows the ratio of observed data to predicted yields.
The dark purple band shows the statistical uncertainty in the background prediction,
while the light blue band shows the total uncertainty.}
\label{fig:newmultilep:yields}
\end{figure}

\begin{figure}[htbp]
\centering
\includegraphics[width=0.45\textwidth]{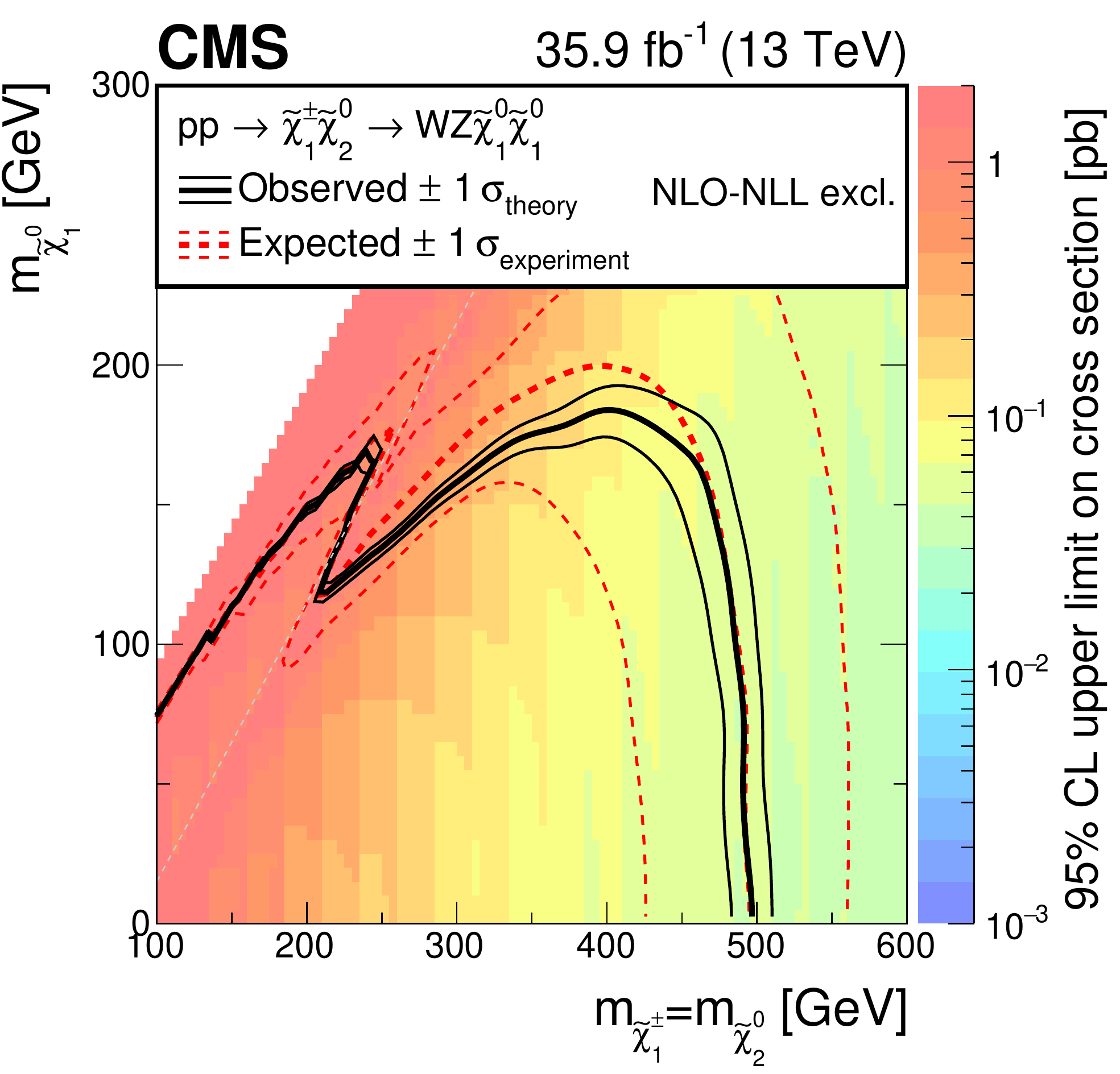}
\caption{
The 95\% confidence level upper limit on the production cross section in the plane of \mfirstcharg and \mfirstchi
for the model of \chargneut production with the \wz topology,
using only the search requiring three or more leptons as described in Section~\ref{sec:newmultilep}.
  The thick solid black (dashed red) curve represents the observed (expected) exclusion contour assuming the theory cross sections.
  The area below each curve is the excluded region.
      The thin dashed red lines indicate the $\pm1\sigma_{\text{experiment}}$ uncertainty.
      The thin black lines show the effect of the theoretical
      uncertainties (${\pm}1\sigma_{\text{theory}}$) on the signal cross section.
      The color scale shows the observed limit at 95\% CL on the signal production cross section.
}
\label{fig:newmultilep:interpretation}
\end{figure}

Distributions of key kinematic observables for the events entering the search
regions are shown in Fig.~\ref{fig:newmultilep:results:kin} with two representative
signal mass points included. 
The data agree with the prediction within systematic uncertainties,
which are dominated at high \MT\ and \ptmiss by the \wz control region statistical precision as described above. 
This uncertainty is taken as correlated across signal region bins.
The comparison between expected and observed
yields in the search regions is shown in Fig.~\ref{fig:newmultilep:yields} and
Table~\ref{tab:newmultilep:yields}.
No significant deviations from the SM expectations are observed.
The predicted background yields and uncertainties presented in this section are 
used as inputs to the likelihood fit for interpretation, described in Section~\ref{sec:interp}.
The interpretation of the results in the \wz topology at 95\% confidence level (CL) is
presented in Fig.~\ref{fig:newmultilep:interpretation}.
Compared to Ref.~\cite{ewkino2016}, the expected lower mass limit in the \wz corridor
has improved from around $(\msecondchi,\mfirstchi) = (200,100)$ to around $(225,125)\GeV$,
while the observed limit has improved by around 60\GeV in both mass values.
The expected limit contour for signal points with $\msecondchi - \mfirstchi > m_{\PZ}$ has also
improved by as much as 25\GeV due to the new selections.
The upper limit on the \chargneut\ production cross section has improved by a factor of 2.

The event selections listed in Table~\ref{tab:newmultilep:categories} are used to replace the selections for category~A in
Ref.~\cite{ewkino2016} in the combination below with other analyses, when interpreting results
in the models with either 100\% or 50\% branching fraction to the SUSY \wz topology.
In this case, the systematic uncertainties in the background prediction are treated as being fully correlated with
the other categories from Ref.~\cite{ewkino2016}.

\begin{table}[tbh]
\centering
\topcaption{Expected and observed event yields in the search regions. For each bin,
the first number corresponds to the expected yield and its total uncertainty
while the second number gives the observation.
Where ranges of values are given for the selections, the lower bound is inclusive while the upper bound is exclusive,
\eg, $75 \leq \mll < 105\GeV$.
}
\label{tab:newmultilep:yields}
\resizebox{0.8\textwidth}{!}{
\begin{tabular}{|c|c|c|cc|cc|cc|}
\hline
\mll\ (\GeVns{}) & \MT\ (\GeVns{}) & \ptmiss (\GeVns{}) & \multicolumn{2}{c|}{$\HT< 100\GeV$} & \multicolumn{2}{c|}{$100 \leq \HT<200\GeV$} & \multicolumn{2}{c|}{$\HT\geq 200\GeV$} \\
\hline\hline
\multirow{11}{*}{0--75  }  & \multirow{4}{*}{0--100}   & 50--100  & \multicolumn{2}{c}{   175 $\pm$     20}  & \multicolumn{2}{c|}{166}  & \multirow{4}{*}{    39 $\pm$      6} & \multirow{4}{*}{41} \\ \cline{3-7}
                            &                               & 100--150 & \multicolumn{2}{c}{    27 $\pm$      4}  & \multicolumn{2}{c|}{23}  &  & \\ \cline{3-7}
                            &                               & 150--200 & \multicolumn{2}{c}{     5 $\pm$      1}  & \multicolumn{2}{c|}{6}  &  & \\ \cline{3-7}
                            &                               & $\geq$200 & \multicolumn{2}{c}{  2.5 $\pm$   0.8}  & \multicolumn{2}{c|}{1}  &  & \\ \cline{2-9}
                            & \multirow{3}{*}{100--160} & 50--100  & \multicolumn{2}{c}{    50 $\pm$      8}  & \multicolumn{2}{c|}{56}  & \multirow{3}{*}{    10 $\pm$      3} & \multirow{3}{*}{13} \\ \cline{3-7}
                            &                               & 100--150 & \multicolumn{2}{c}{    12 $\pm$      3}  & \multicolumn{2}{c|}{13}  & & \\ \cline{3-7}
                            &                               & $\geq$150 & \multicolumn{2}{c}{  1.2 $\pm$   0.4}  & \multicolumn{2}{c|}{1}  & & \\ \cline{2-9}
                            & \multirow{4}{*}{$\geq$160}     & 50--100  & \multicolumn{2}{c}{    12 $\pm$      2}  & \multicolumn{2}{c|}{13}  & \multirow{4}{*}{     6 $\pm$      2} & \multirow{4}{*}{11} \\ \cline{3-7}
                            &                               & 100--150 & \multicolumn{2}{c}{    11 $\pm$      3}  & \multicolumn{2}{c|}{14}  & &  \\ \cline{3-7}
                            &                               & 150--200 & \multicolumn{2}{c}{  2.6 $\pm$   0.9} & \multicolumn{2}{c|}{2} & &  \\ \cline{3-7}
                            &                               & $\geq$200 & \multicolumn{2}{c}{  1.2 $\pm$   0.5} & \multicolumn{2}{c|}{1} & &  \\ \hline
\multirow{18}{*}{75--105}  & \multirow{6}{*}{0--100}   & 50--100  & \multicolumn{2}{c|}{(WZ CR)}                 &    279 $\pm$     34 & 250 & \multirow{2}{*}{   310 $\pm$     40} & \multirow{2}{*}{292} \\ \cline{3-7}
                            &                               & 100--150 &    286 $\pm$     44 & 260 &     87 $\pm$     13 & 81 &       &  \\ \cline{3-9}
                            &                               & 150--200 &     62 $\pm$     14 & 51 &     26 $\pm$      6 & 20 & \multirow{2}{*}{    81 $\pm$     18} & \multirow{2}{*}{69} \\ \cline{3-7}
                            &                               & 200--250 &     20 $\pm$      5 & 10 &      8 $\pm$      2 & 10 &       &  \\ \cline{3-9}
                            &                               & 250--350 & \multirow{2}{*}{    16 $\pm$      4} & \multirow{2}{*}{9} & \multirow{2}{*}{     6 $\pm$      1} & \multirow{2}{*}{5}&     25 $\pm$      6 & 23 \\ \cline{3-3}\cline{8-9}
                            &                               & $\geq$350 &                       &   &  &       &     13 $\pm$      3 & 8 \\ \cline{2-9}
                            & \multirow{6}{*}{100--160} & 50--100  &    321 $\pm$     42 & 297   &     54 $\pm$      8 & 49 &     45 $\pm$      6 & 45 \\ \cline{3-9}
                            &                               & 100--150 &     50 $\pm$     14 & 38   &     11 $\pm$      3 & 11 &     14 $\pm$      3 & 12 \\ \cline{3-9}
                            &                               & 150--200 &      5 $\pm$      2 & 2   &   2.2 $\pm$   0.9 & 2 &      4 $\pm$      2 & 5 \\ \cline{3-9}
                            &                               & 200--250 & \multirow{3}{*}{  1.1 $\pm$   0.5} & \multirow{3}{*}{2}& \multirow{3}{*}{  0.5 $\pm$   0.4} & \multirow{3}{*}{2} &   1.9 $\pm$   0.8 & 1 \\ \cline{3-3}\cline{8-9}
                            &                               & 250--300 &          &              &   &   &   1.8 $\pm$   0.8 & 2 \\ \cline{3-3}\cline{8-9}
                            &                               & $\geq$300 &          &              &   &   &   1.0 $\pm$   0.5 & 1 \\ \cline{2-9}
                            & \multirow{6}{*}{$\geq$160}     & 50--100  &     25 $\pm$      6 & 18     &      6 $\pm$      2 & 5 &      9 $\pm$      3 & 12 \\ \cline{3-9}
                            &                               & 100--150 &     12 $\pm$      5 & 13     &   3.0 $\pm$   1.3 & 2 &      4 $\pm$      2 & 2 \\ \cline{3-9}
                            &                               & 150--200 &      5 $\pm$      2 & 5     &   1.1 $\pm$   0.4 & 0 &   2.0 $\pm$   0.7 & 2 \\ \cline{3-9}
                            &                               & 200--250 & \multirow{3}{*}{     4 $\pm$      2} & \multirow{3}{*}{2} & \multirow{3}{*}{  0.9 $\pm$   0.4} & \multirow{3}{*}{3} &   1.5 $\pm$   0.7 & 2 \\ \cline{3-3}\cline{8-9}
                            &                               & 250--300 &                    &   &      &   &   0.6 $\pm$   0.3 & 1 \\ \cline{3-3}\cline{8-9}
                            &                               & $\geq$300 &                    &   &      &   &   1.1 $\pm$   0.5 & 1 \\ \hline

\multirow{ 3}{*}{$\geq$105   } & 0--100                    & $\geq$50  & \multicolumn{3}{c}{   173 $\pm$     21} & \multicolumn{3}{c|}{170} \\ \cline{2-9}
                            & 100--160                  & $\geq$50  & \multicolumn{3}{c}{    44 $\pm$      7} & \multicolumn{3}{c|}{28} \\ \cline{2-9}
                            & $\geq$160                     & $\geq$50  & \multicolumn{3}{c}{    23 $\pm$      6} & \multicolumn{3}{c|}{12} \\ \hline
\end{tabular}
}
\end{table}

\section{Interpretation}
\label{sec:interp}

The results of the searches described in Sections~\ref{sec:searches} and~\ref{sec:newmultilep}
are interpreted using the simplified models introduced in Section~\ref{sec:sigs}.
Cross section limits as a function of the SUSY particle masses are set using a modified
frequentist approach, employing the CL$_{\text{s}}$ criterion and an asymptotic
formulation~\cite{Junk:1999kv,Read:2002hq,Cowan:2010js, ATL-PHYS-PUB-2011-011}.
The uncertainties in the signal efficiency and acceptance and in the background predictions
are incorporated as nuisance parameters. The observed data yields in control regions
are typically incorporated either by a simultaneous maximum likelihood fit of the signal
and control regions or through parameterization using the gamma function. Other nuisance parameters are
implemented using lognormal functions, whose widths reflect the size of the systematic uncertainty,
or as alternate shapes of the relevant distributions.
Within each signal model, the experimental and theoretical uncertainties affecting the signal prediction 
are treated as fully correlated for all analyses.
The dominant uncertainties in the background predictions are not correlated among analyses as they tend to be
either statistical in nature, arising from independent control regions, 
or uncertainties in the prediction methods, which are unique to each analysis.
For each signal topology, the analyses with a check mark in Table~\ref{tab:searches_models} are combined to place exclusion limits.

The following sources of uncertainty in the signal acceptance and efficiency are assumed to be fully correlated among analyses:
determination of the integrated luminosity, lepton identification and isolation efficiency,
lepton efficiency modeling in fast simulation, b tagging efficiency, jet energy scale,
modeling of \ptmiss in fast simulation,
modeling of ISR, simulation of pileup,
and variations of the generator factorization and renormalization scales.
Variations in the PDF set used are found to primarily affect the signal acceptance by changing 
the \pt\ distribution of the initially-produced sparticle pair, $\firstcharg\secondchi$ or $\firstchi\firstchi$.
This is already incorporated in the empirical uncertainty in the modeling of ISR as described in Section~\ref{sec:recosamples},
and we therefore do not apply a dedicated uncertainty in signal acceptance from PDF variations.
All analyses also include the statistical uncertainty of the simulated signal samples, which is taken as being uncorrelated in every bin,
and the uncertainty in the modeling of the trigger efficiency, which is also taken as uncorrelated given the different
trigger requirements applied in each analysis.
Some analyses have additional uncertainties beyond these, such as the uncertainty in the modeling of the diphoton mass resolution
for the \hgg\ analysis, which are analysis-specific and treated as being uncorrelated.

\begin{figure}[htbp]
\centering
\includegraphics[width=0.6\textwidth]{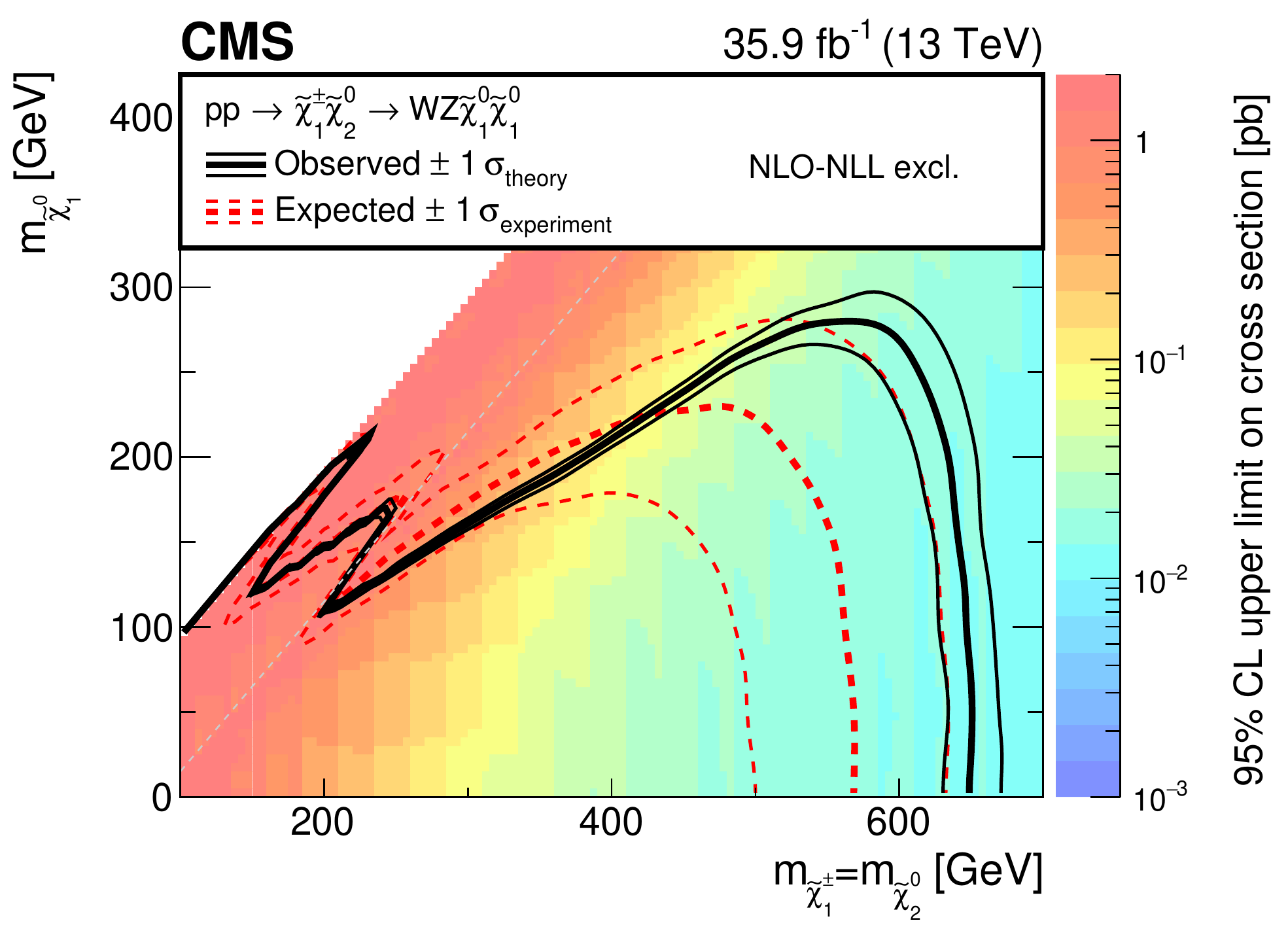}
\includegraphics[width=0.6\textwidth]{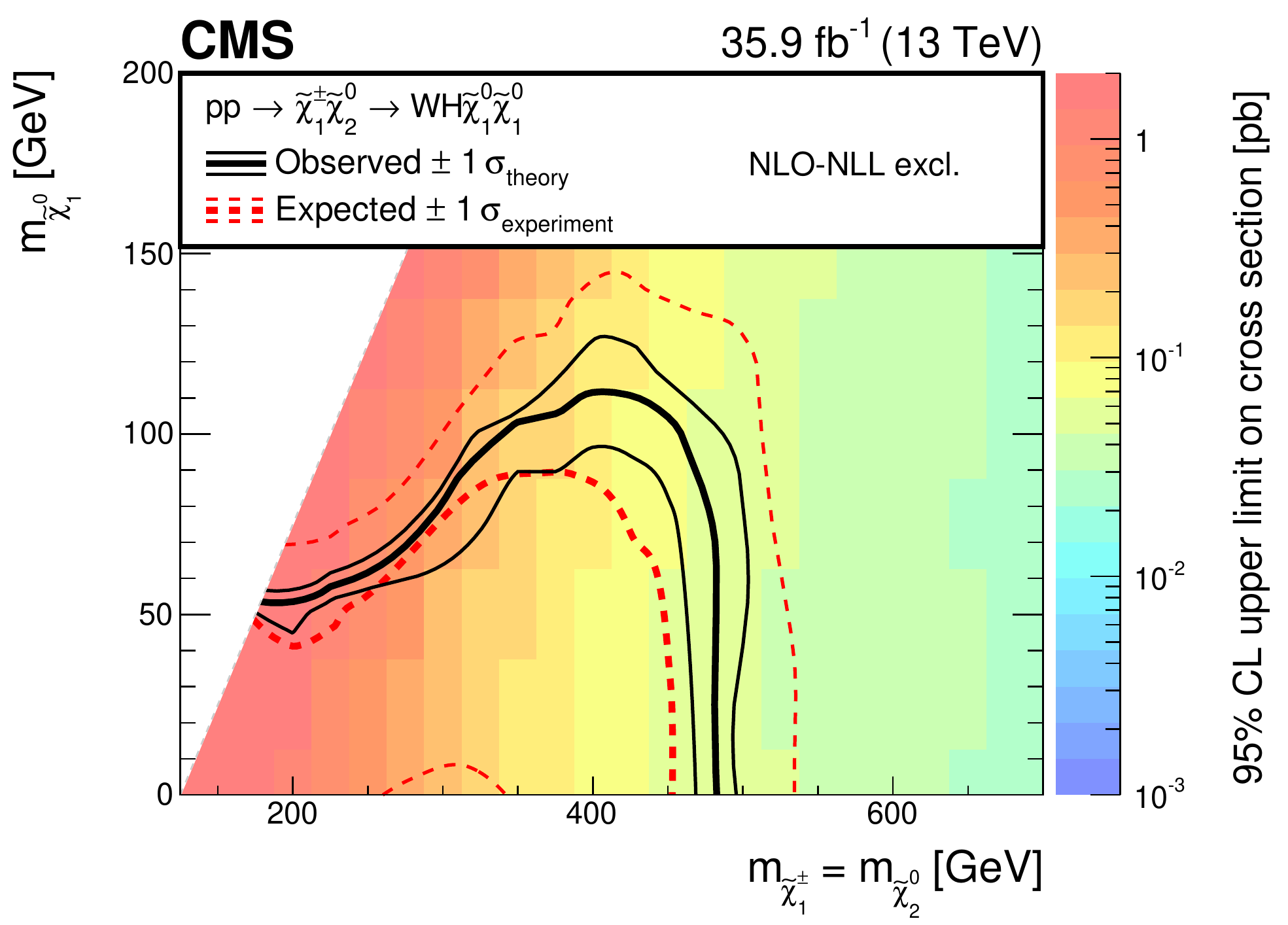}
\includegraphics[width=0.6\textwidth]{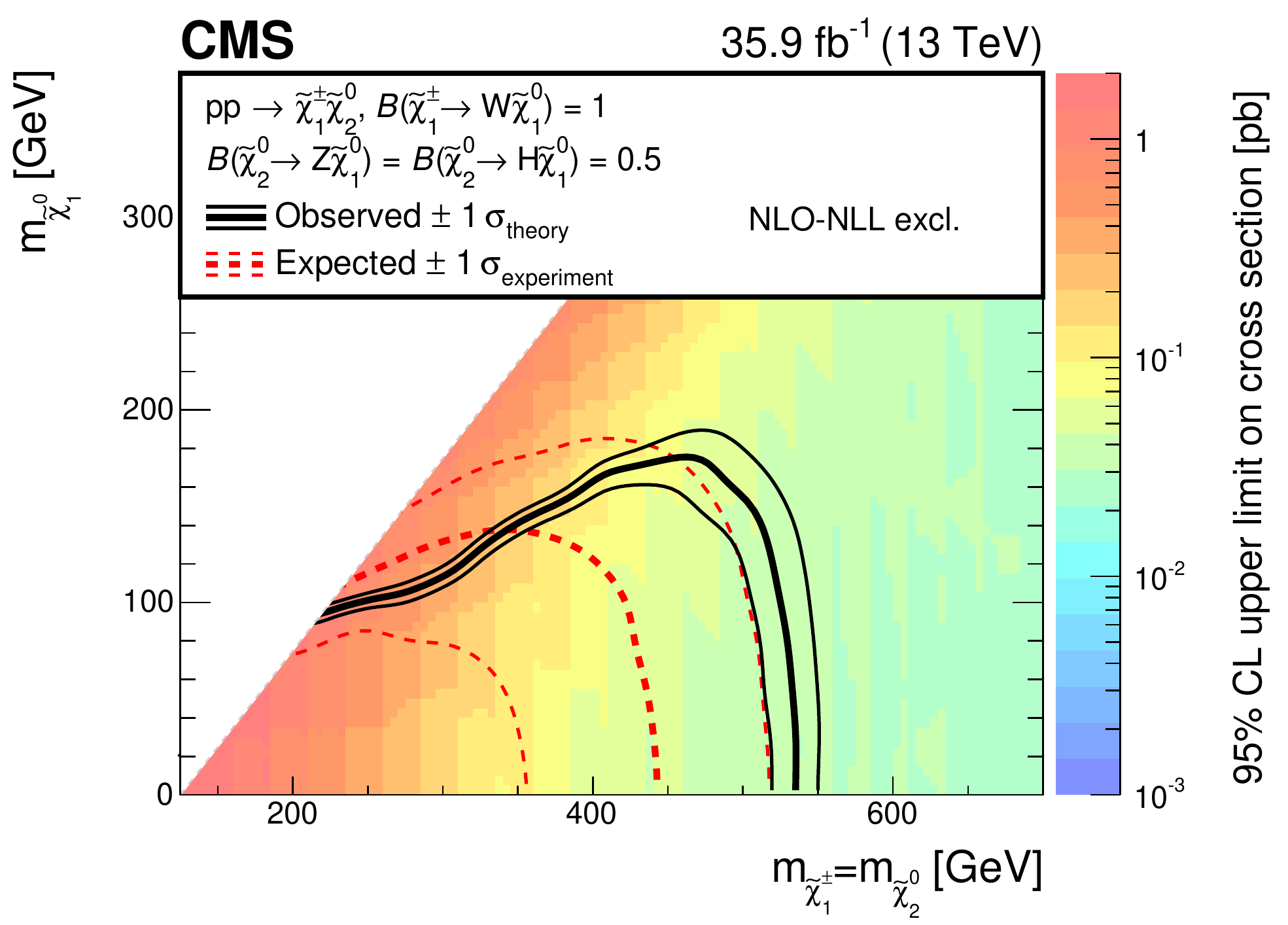}
\caption{The 95\% CL upper limits on the production cross sections in the plane of \mfirstcharg and \mfirstchi
  for the models of \chargneut production with (upper) the \wz topology, (middle) the \wh topology,
  or (lower) the mixed topology with 50\% branching fraction to each of \wz and \wh.
  The thick solid black (dashed red) curve represents the observed (expected) exclusion contour assuming the theory cross sections.
  The area below each curve is the excluded region.
      The thin dashed red lines indicate the $\pm1\sigma_{\text{experiment}}$ uncertainty.
      The thin black lines show the effect of the theoretical
      uncertainties ($\pm1\sigma_{\text{theory}}$) on the signal cross section.
      The color scale shows the observed limit at 95\% CL on the signal production cross section.
}
\label{fig:limits_c1n2}
\end{figure}

\begin{figure}[htbp]
\centering
\includegraphics[width=0.6\textwidth]{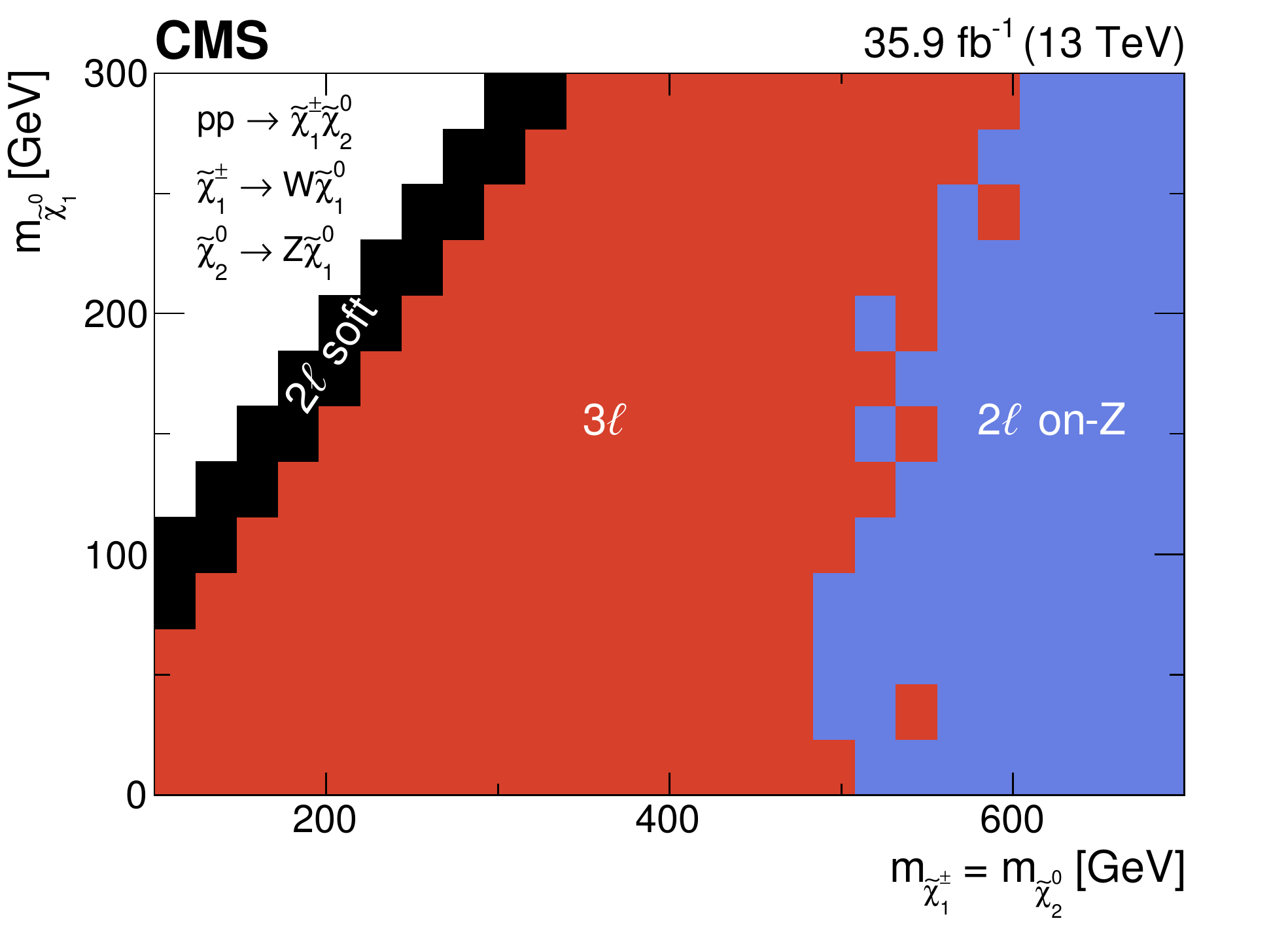}
\includegraphics[width=0.6\textwidth]{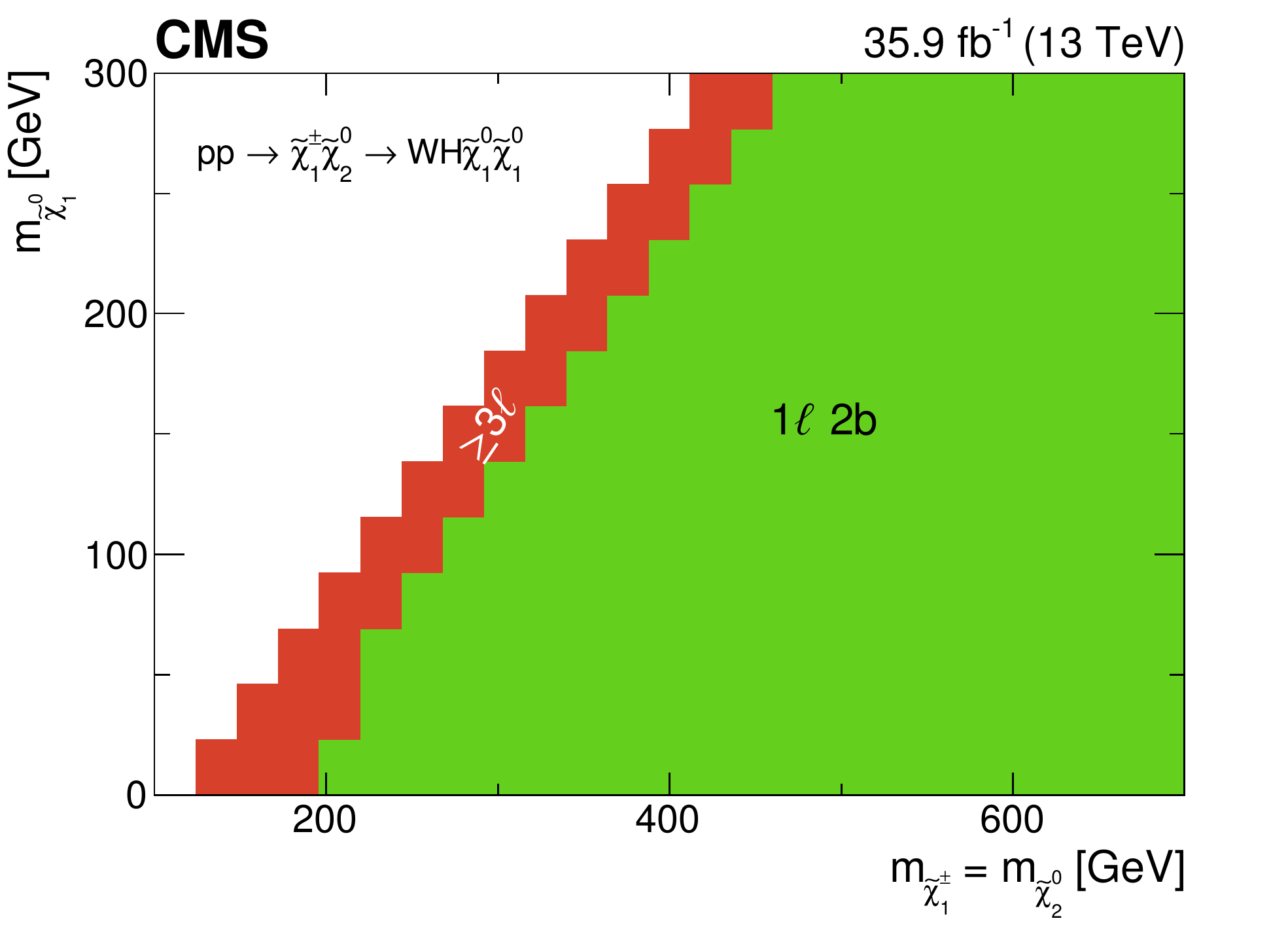}
\includegraphics[width=0.6\textwidth]{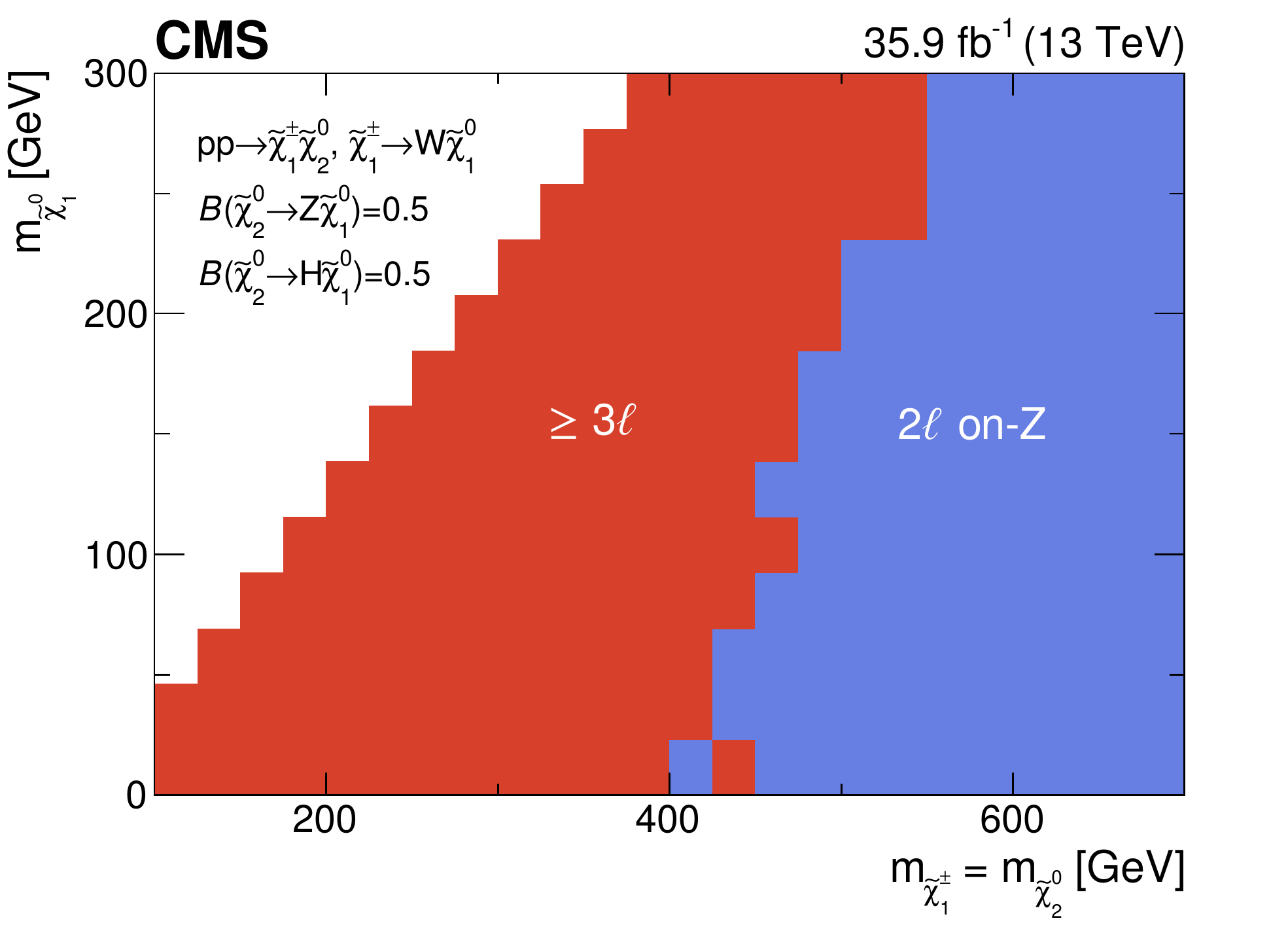}
\caption{The analysis with the best expected exclusion limit at each point in the plane of \mfirstcharg and \mfirstchi
  for the models of \chargneut production with (upper) the \wz topology, (middle) the \wh topology,
   and (lower) the mixed topology 50\% branching fraction to each of \wz and \wh.
}
\label{fig:analyses_c1n2_2d}
\end{figure}

\begin{figure}[htbp]
\centering
\includegraphics[width=0.45\textwidth]{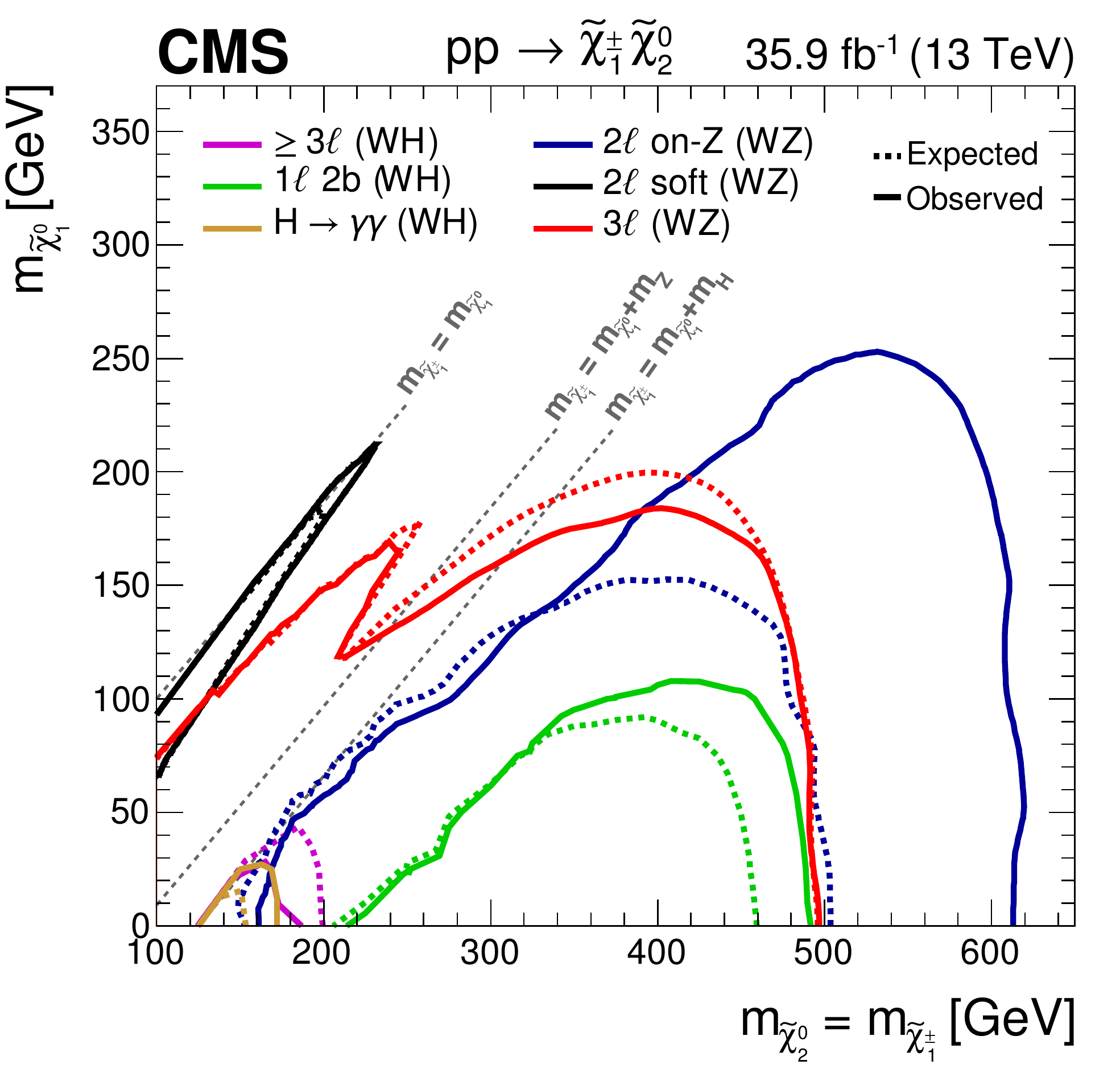}
\includegraphics[width=0.45\textwidth]{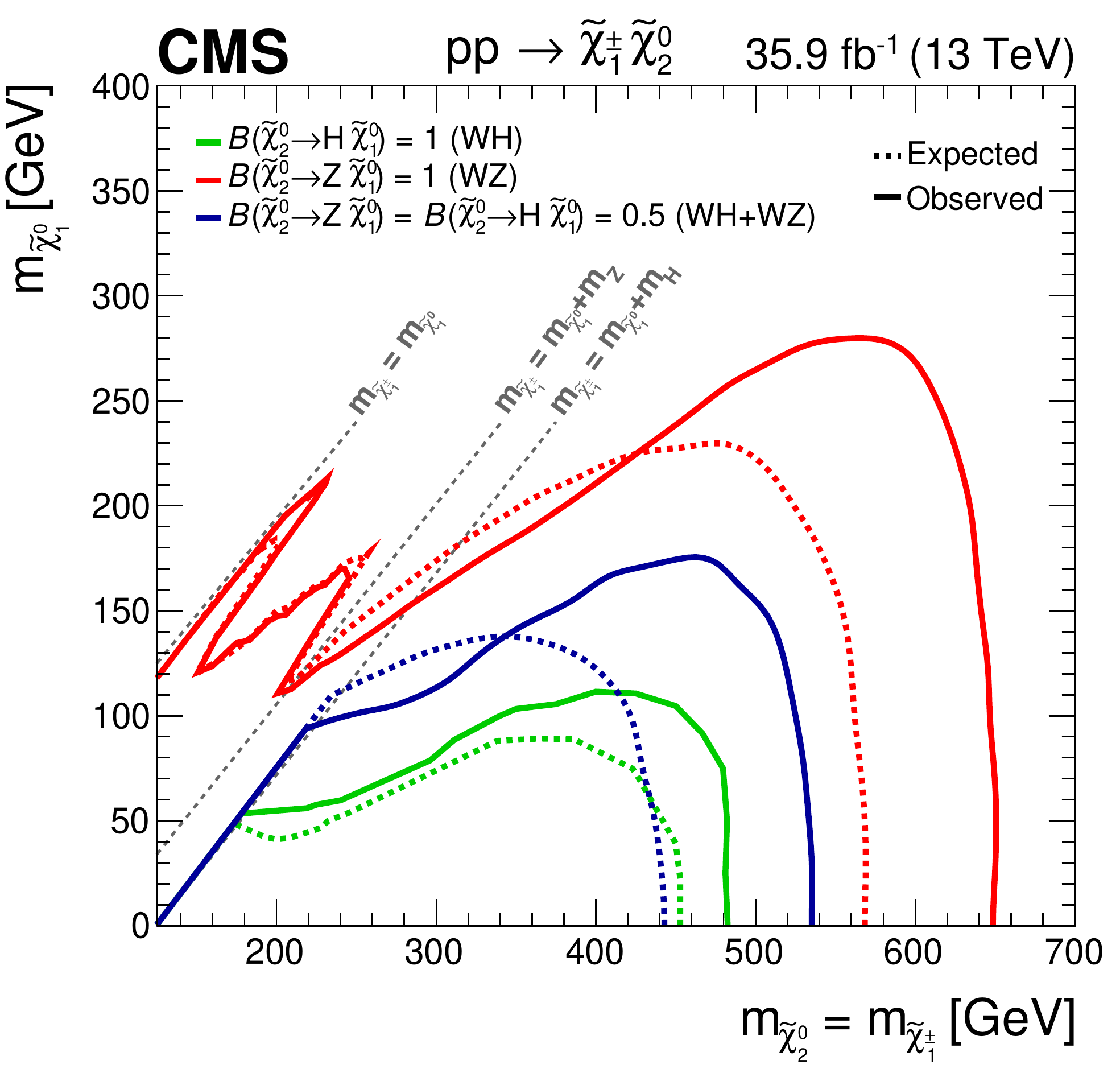}
\caption{Exclusion contours at 95\% CL in the plane of \mfirstcharg and \mfirstchi
  for the models of \chargneut production (left) for the individual analyses
  and (right) for the combination of analyses.
  The decay modes assumed for each contour are given in the legends.
}
\label{fig:limits_c1n2_summary}
\end{figure}

For the models of \chargneut production, 95\% confidence level exclusion limits are presented
in the plane of \mfirstcharg and \mfirstchi.
Figure~\ref{fig:limits_c1n2} shows the exclusion limits for the combination of analyses
for the \wz topology, the \wh topology, and the mixed topology with 50\% branching fraction to each of the \wz and \wh channels.
Figure~\ref{fig:analyses_c1n2_2d} shows the analysis with the best expected limit for each point in the plane
for the same topologies.  The on-\PZ dilepton analysis generally gives the best sensitivity for large values of
$\Delta m = \msecondchi - \mfirstchi$.  The search for three light-flavor leptons provides the best sensitivity
at intermediate values of $\Delta m$, including the region where $\Delta m \approx m_{\PZ}$,
while the soft-dilepton analysis provides unique sensitivity to the smallest values of $\Delta m$.
Figure~\ref{fig:limits_c1n2_summary} (left) shows the observed and expected limit contours for each of the
individual analyses considered in the combination, and Fig.~\ref{fig:limits_c1n2_summary} (right)
shows the results from the combination for all three topologies considered.
For a massless LSP \firstchi, the combined result gives an observed (expected) limit in \mfirstcharg\
of about 650\,(570)\GeV for the \wz topology, 480\,(455)\GeV for the \wh topology, and 535\,(440)\GeV for the mixed topology.
The combination also excludes intermediate mass values that were not excluded by individual analyses,
including \mfirstcharg values between 180 and 240\GeV for a massless LSP in the \wh topology.

\begin{figure}[htbp]
\centering
\includegraphics[width=0.65\textwidth]{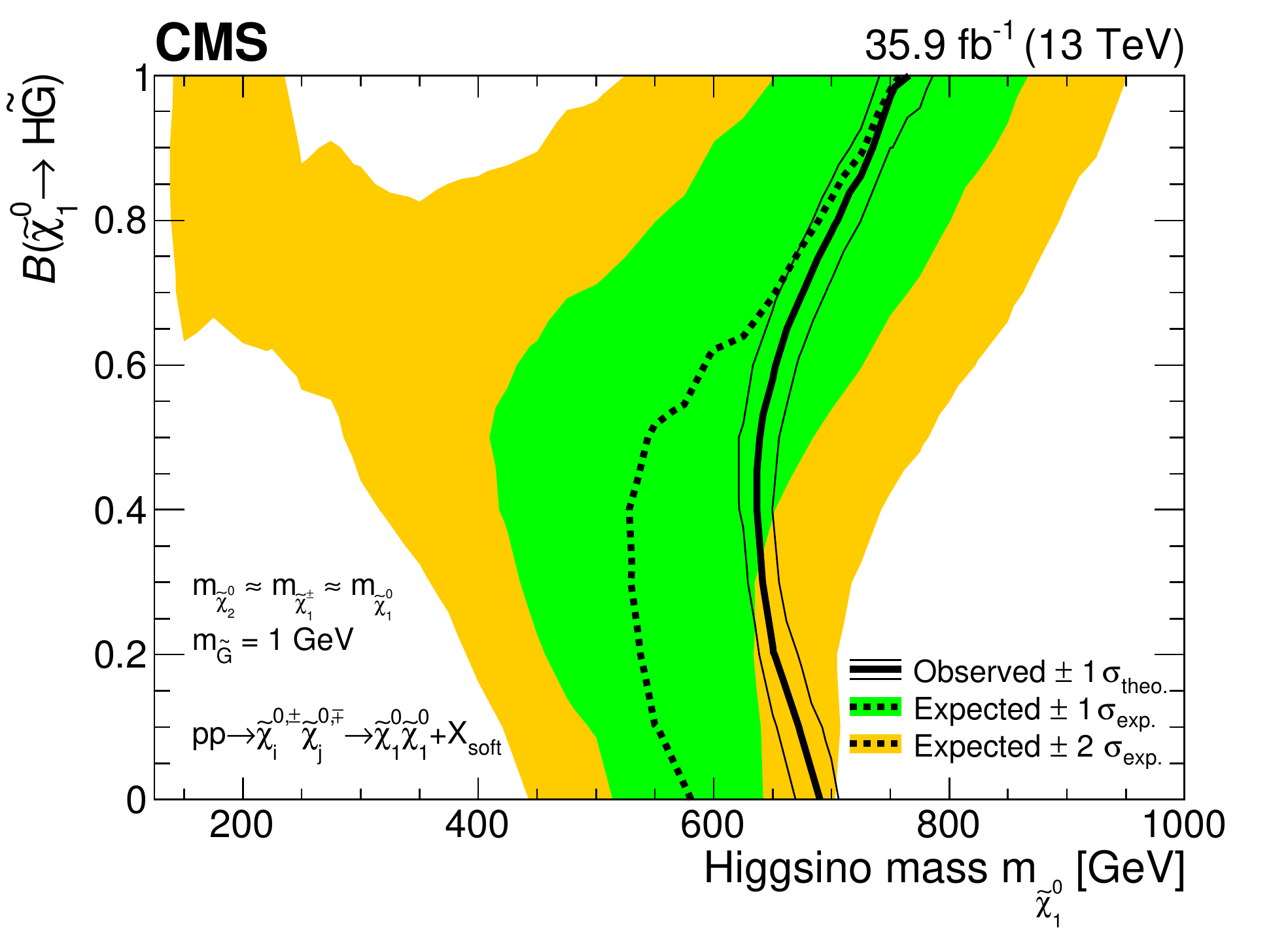}
\caption{Combined exclusion contours at the 95\% CL in the plane of \mfirstchi\ and \bfchih\
  for the model of \neutneut\ production.
  The area to the left of or below the solid (dashed) black curve represents the observed (expected) exclusion region.
      The green and yellow bands indicate the
      $\pm$1 and 2$\sigma$ uncertainties in the expected limit.
      The thin black lines show the effect of the theoretical
      uncertainties (${\pm}1\sigma_{\text{theory}}$) on the signal cross section.
}
\label{fig:limits_n1n1_2d}
\end{figure}

For the models of \neutneut production, the exclusion limits are presented
in the plane of \mfirstchi and the branching fraction \bfchih.
The decay $\firstchi \to \PZ\gravitino$ is assumed to make up the remainder of the branching fraction.
Figure~\ref{fig:limits_n1n1_2d} shows the observed and expected limits from the combination in this plane.
The expected mass exclusion limit varies between about 550 and 750\GeV,
being least stringent around $\bfchih = 0.4$.
The observed limit ranges between about 650 and 750\GeV, allowing us to exclude masses below 650\GeV independent of this branching fraction.

\begin{figure}[htbp]
\centering
\includegraphics[width=0.65\textwidth]{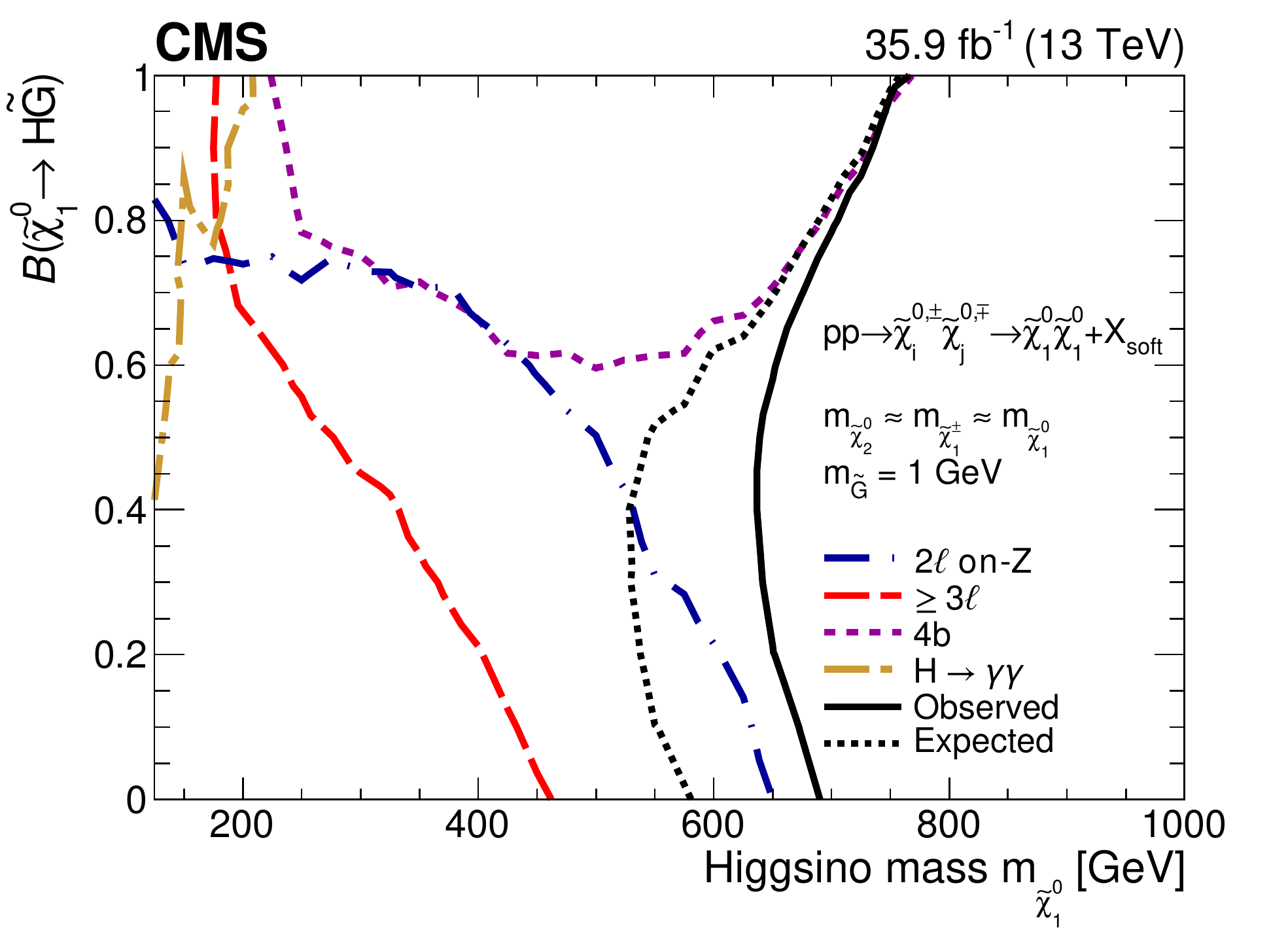}
\caption{Observed exclusion contours at the 95\% CL in the plane of \mfirstchi\ and \bfchih\
  for the model of \neutneut\ production for each individual analysis compared with the combination.
  For the 4\PQb\ contour, the region above is excluded, while for all others,
  the region to the left is excluded.
  The 4\PQb\ search drives the exclusion at large values of \bfchih while the on-\PZ dilepton
  and multilepton searches are competing at lower values of \bfchih.
}
\label{fig:indiv_limits_n1n1_2d}
\end{figure}

\begin{figure}[htbp]
\centering
\includegraphics[width=0.6\textwidth]{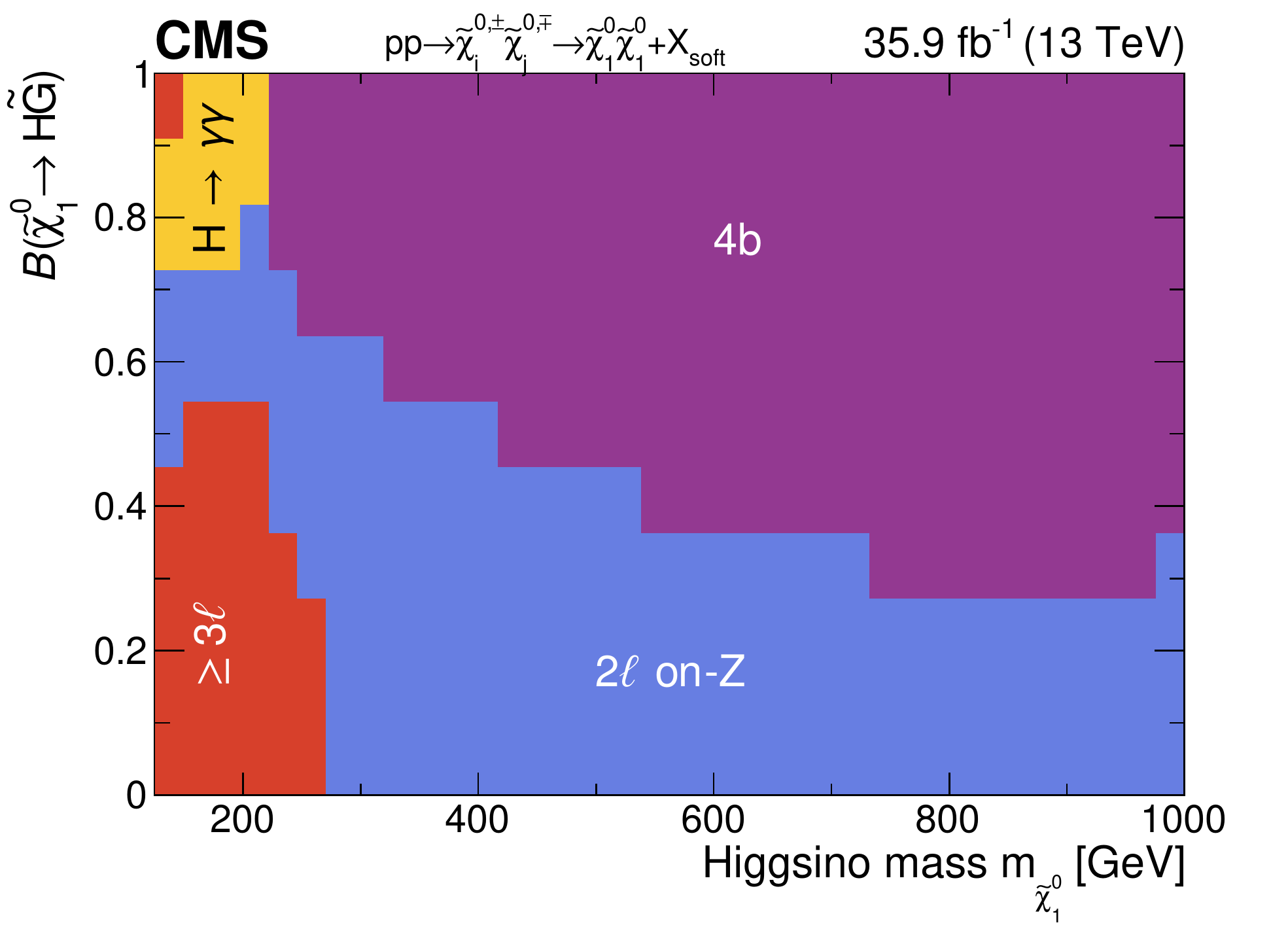}
\caption{The analysis with the best expected exclusion limit at each point in the plane of \mfirstchi and \bfchih\
  for the model of \neutneut production.
}
\label{fig:analyses_n1n1_2d}
\end{figure}

\begin{figure}[htbp]
\centering
\includegraphics[width=0.6\textwidth]{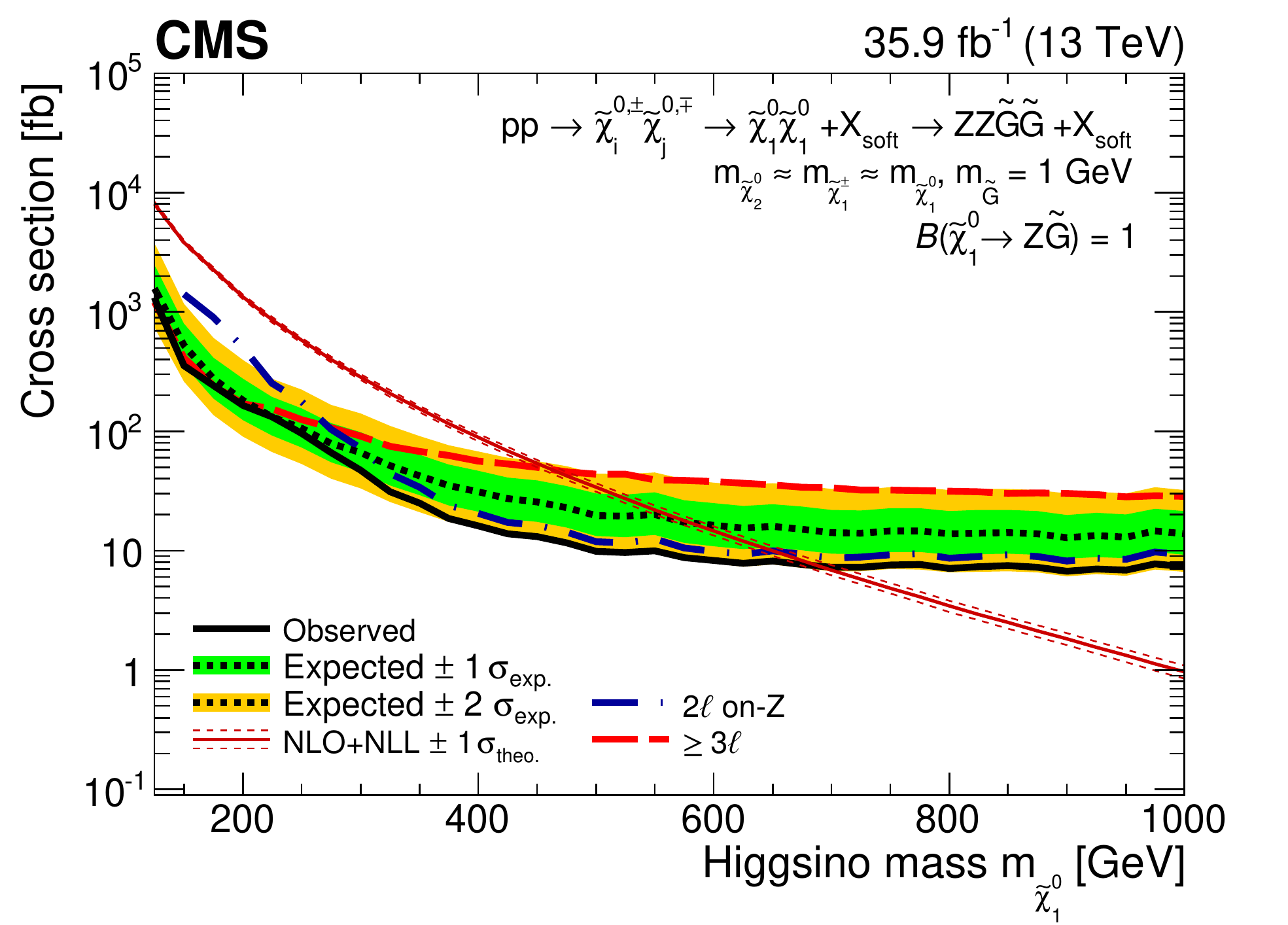}
\includegraphics[width=0.6\textwidth]{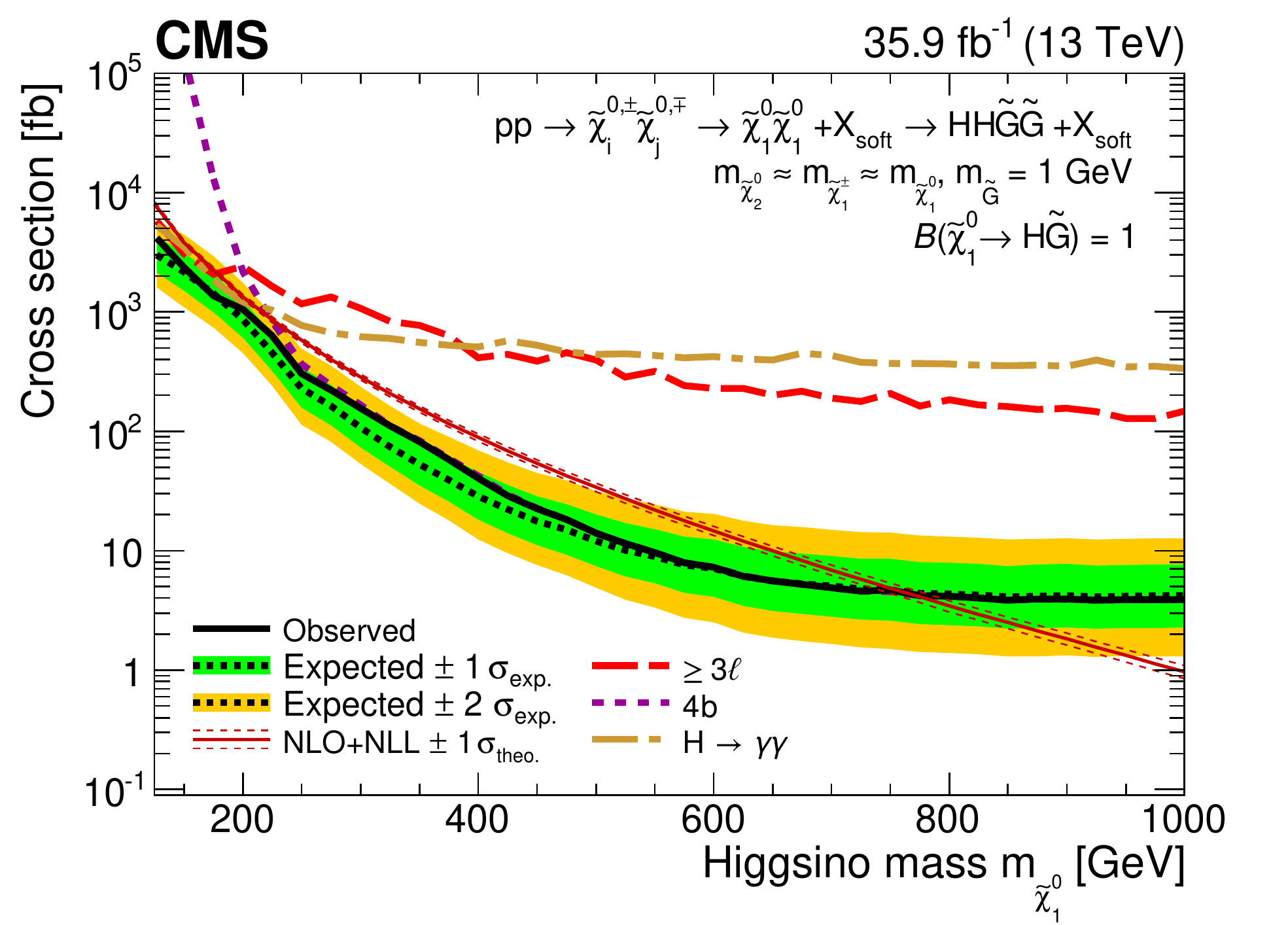}
\includegraphics[width=0.6\textwidth]{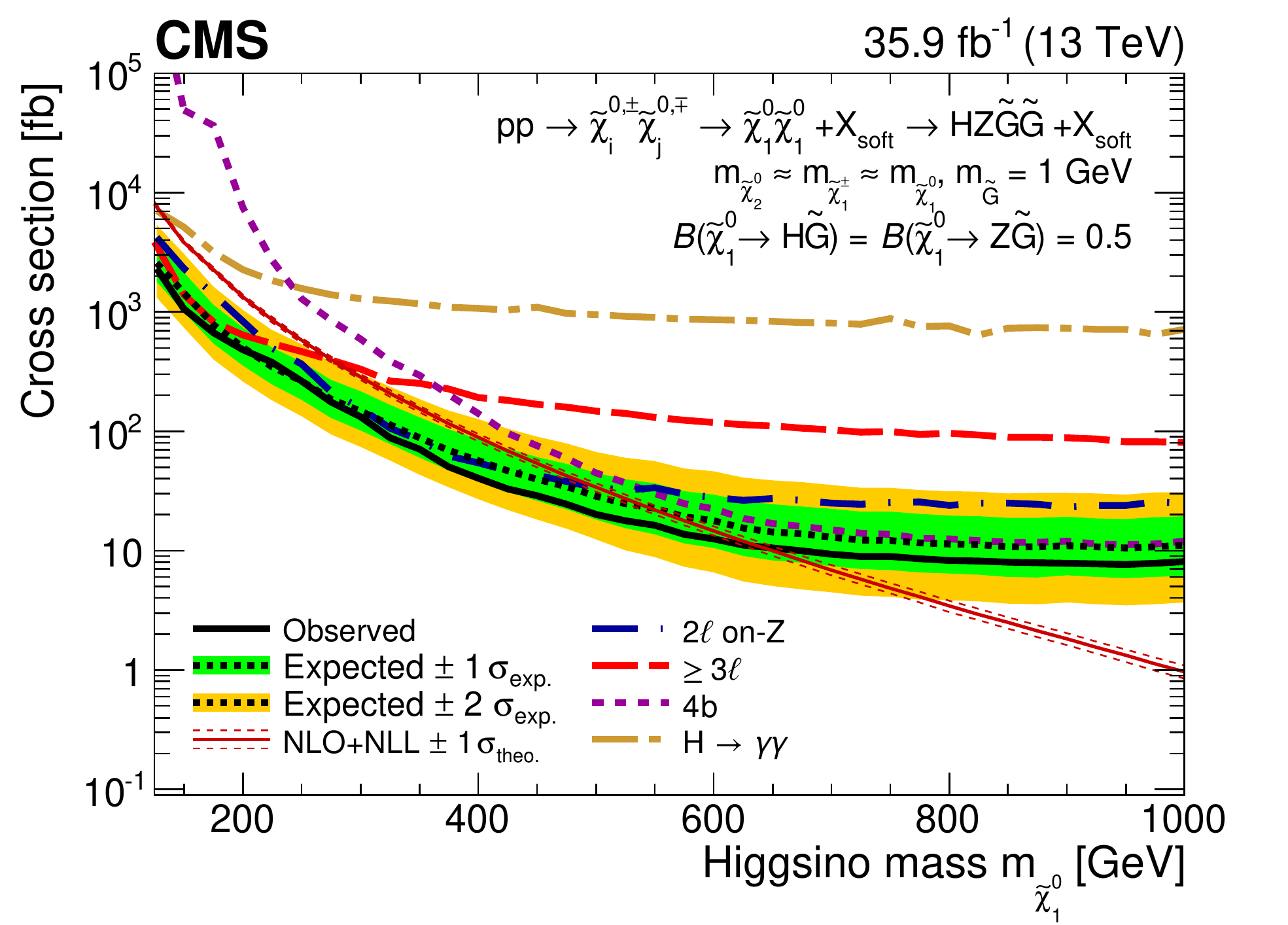}
\caption{The 95\% CL upper limits on the production cross sections as a function of \mfirstchi\
  for the model of \neutneut\ production with
three choices of \bfchih: (upper) 0\%, yielding the \zz topology,
(middle) 100\%, yielding the \hh topology, and (lower) 50\%, yielding the \zh mixed topology.
      The solid black line represents the observed exclusion.  The dashed black line represents
      the expected exclusion, while the green and yellow bands indicate the
      $\pm$1 and 2$\sigma$ uncertainties in the expected limit.
      The red line shows the theoretical cross section with its uncertainty.
      The other lines in each plot show the observed exclusion for individual analyses.
}
\label{fig:limits_n1n1_1d}
\end{figure}

Figure~\ref{fig:indiv_limits_n1n1_2d} shows the observed limits from each analysis separately compared with the combined result.
Figure~\ref{fig:analyses_n1n1_2d} shows the analysis with the best expected exclusion limit for each point in the same plane.
At higher values of \mfirstchi, the searches for at least one hadronically decaying boson provide the best sensitivity, the 4\PQb\ search
when \bfchih is large and the on-\PZ dilepton search when it is smaller.
At lower values of \mfirstchi, below around 200\GeV, the \hgg analysis is most sensitive when \bfchih is large,
while the three or more lepton search is dominant when it is small.
Figure~\ref{fig:limits_n1n1_1d} then shows the exclusion limits as a function of \mfirstchi for
three choices of \bfchih: 0\%, yielding the \zz topology;
100\%, yielding the \hh topology; and 50\%, yielding a mix of events from the \zz, \hh, and \zh topologies.

\section{Summary}

A number of searches for the electroweak production of charginos and neutralinos predicted in supersymmetry (SUSY)
have been performed in different final states.
All searches considered here use proton-proton collision data at $\sqrt{s}=13\TeV$,
recorded with the CMS detector at the LHC and corresponding to an integrated luminosity of $\fullLumi$.
No significant deviations from the standard model expectations have been observed.

A targeted search requiring three or more charged leptons (electrons or muons) has been presented,
focusing on chargino-neutralino production where the difference in mass between \secondchi\ and \firstchi\
is approximately equal to the mass of the \PZ{} boson,
and no significant deviations from the standard model predictions are observed.
This search is interpreted in a simplified model scenario of SUSY
chargino-neutralino ($\firstcharg\secondchi$) production with decays $\firstcharg\to\PW^{\pm}\firstchi$ and $\secondchi\to\PZ\firstchi$,
where \firstchi\ is the lightest SUSY particle (LSP).
In the targeted phase space, the expected and observed 95\% confidence level exclusion limits
extend to 225\GeV in the mass of \secondchi and 125\GeV in the mass of \firstchi,
improving the observed limits from the previous publication by up to 60\GeV~\cite{ewkino2016}.

A statistical combination of several searches is performed and interpreted in the context of simplified models of either
chargino-neutralino production, or neutralino pair production in a gauge-mediated SUSY breaking (GMSB) scenario.
For a massless LSP \firstchi in the chargino-neutralino model, the combined result gives an observed (expected) limit in the \firstcharg mass
of about 650\,(570)\GeV for the \wz topology, 480\,(455)\GeV for the \wh topology, and 535\,(440)\GeV for the mixed topology.
Compared to the results of individual analyses, the combination improves the observed exclusion limit by up to 40\GeV
in the masses of \firstcharg and \secondchi in the chargino-neutralino model.
The combination also excludes intermediate mass values that were not excluded by individual analyses,
including \firstcharg masses between 180 and 240\GeV in the \wh topology.
In the GMSB neutralino pair model, the combined result gives an observed (expected) limit in the \firstchi mass of 650--750 (550--750)\GeV.
The combined result improves the observed limit by up to 200\GeV in the mass of \firstchi in the GMSB neutralino pair model,
depending on the branching fractions for the SUSY particle decays.
These results represent the most stringent constraints to date for all models considered.

\begin{acknowledgments}
We congratulate our colleagues in the CERN accelerator departments for the excellent performance of the LHC and thank the technical and administrative staffs at CERN and at other CMS institutes for their contributions to the success of the CMS effort. In addition, we gratefully acknowledge the computing centers and personnel of the Worldwide LHC Computing Grid for delivering so effectively the computing infrastructure essential to our analyses. Finally, we acknowledge the enduring support for the construction and operation of the LHC and the CMS detector provided by the following funding agencies: BMWFW and FWF (Austria); FNRS and FWO (Belgium); CNPq, CAPES, FAPERJ, and FAPESP (Brazil); MES (Bulgaria); CERN; CAS, MoST, and NSFC (China); COLCIENCIAS (Colombia); MSES and CSF (Croatia); RPF (Cyprus); SENESCYT (Ecuador); MoER, ERC IUT, and ERDF (Estonia); Academy of Finland, MEC, and HIP (Finland); CEA and CNRS/IN2P3 (France); BMBF, DFG, and HGF (Germany); GSRT (Greece); OTKA and NIH (Hungary); DAE and DST (India); IPM (Iran); SFI (Ireland); INFN (Italy); MSIP and NRF (Republic of Korea); LAS (Lithuania); MOE and UM (Malaysia); BUAP, CINVESTAV, CONACYT, LNS, SEP, and UASLP-FAI (Mexico); MBIE (New Zealand); PAEC (Pakistan); MSHE and NSC (Poland); FCT (Portugal); JINR (Dubna); MON, RosAtom, RAS, RFBR and RAEP (Russia); MESTD (Serbia); SEIDI, CPAN, PCTI and FEDER (Spain); Swiss Funding Agencies (Switzerland); MST (Taipei); ThEPCenter, IPST, STAR, and NSTDA (Thailand); TUBITAK and TAEK (Turkey); NASU and SFFR (Ukraine); STFC (United Kingdom); DOE and NSF (USA).

\hyphenation{Rachada-pisek} Individuals have received support from the Marie-Curie program and the European Research Council and Horizon 2020 Grant, contract No. 675440 (European Union); the Leventis Foundation; the A. P. Sloan Foundation; the Alexander von Humboldt Foundation; the Belgian Federal Science Policy Office; the Fonds pour la Formation \`a la Recherche dans l'Industrie et dans l'Agriculture (FRIA-Belgium); the Agentschap voor Innovatie door Wetenschap en Technologie (IWT-Belgium); the Ministry of Education, Youth and Sports (MEYS) of the Czech Republic; the Council of Science and Industrial Research, India; the HOMING PLUS program of the Foundation for Polish Science, cofinanced from European Union, Regional Development Fund, the Mobility Plus program of the Ministry of Science and Higher Education, the National Science Center (Poland), contracts Harmonia 2014/14/M/ST2/00428, Opus 2014/13/B/ST2/02543, 2014/15/B/ST2/03998, and 2015/19/B/ST2/02861, Sonata-bis 2012/07/E/ST2/01406; the National Priorities Research Program by Qatar National Research Fund; the Programa Severo Ochoa del Principado de Asturias; the Thalis and Aristeia programs cofinanced by EU-ESF and the Greek NSRF; the Rachadapisek Sompot Fund for Postdoctoral Fellowship, Chulalongkorn University and the Chulalongkorn Academic into Its 2nd Century Project Advancement Project (Thailand); the Welch Foundation, contract C-1845; and the Weston Havens Foundation (USA).
\end{acknowledgments}

\bibliography{auto_generated}

\cleardoublepage \appendix\section{The CMS Collaboration \label{app:collab}}\begin{sloppypar}\hyphenpenalty=5000\widowpenalty=500\clubpenalty=5000\textbf{Yerevan Physics Institute,  Yerevan,  Armenia}\\*[0pt]
A.M.~Sirunyan, A.~Tumasyan
\vskip\cmsinstskip
\textbf{Institut f\"{u}r Hochenergiephysik,  Wien,  Austria}\\*[0pt]
W.~Adam, F.~Ambrogi, E.~Asilar, T.~Bergauer, J.~Brandstetter, E.~Brondolin, M.~Dragicevic, J.~Er\"{o}, M.~Flechl, M.~Friedl, R.~Fr\"{u}hwirth\cmsAuthorMark{1}, V.M.~Ghete, J.~Grossmann, J.~Hrubec, M.~Jeitler\cmsAuthorMark{1}, A.~K\"{o}nig, N.~Krammer, I.~Kr\"{a}tschmer, D.~Liko, T.~Madlener, I.~Mikulec, E.~Pree, N.~Rad, H.~Rohringer, J.~Schieck\cmsAuthorMark{1}, R.~Sch\"{o}fbeck, M.~Spanring, D.~Spitzbart, W.~Waltenberger, J.~Wittmann, C.-E.~Wulz\cmsAuthorMark{1}, M.~Zarucki
% ...
\vskip\cmsinstskip
% ...
1:~~Also at Vienna University of Technology, Vienna, Austria\\
% ...

\end{sloppypar}
\end{document}